\documentclass[runningheads]{llncs}

\usepackage{amsmath}
\usepackage{amssymb}
\usepackage[inline]{enumitem}
\usepackage{graphicx}
\usepackage{listings}
\usepackage{xspace}
\usepackage{algorithm}
\usepackage[noend]{algpseudocode}
\usepackage{subcaption}
\usepackage[table,x11names]{xcolor}
\usepackage{mathtools}
\usepackage[createShortEnv, conf={restate, text proof={}}]{proof-at-the-end}
\usepackage{hyperref}
\usepackage{caption}
\usepackage{tabularx}
\usepackage{booktabs}
\usepackage{makecell}
\usepackage{stmaryrd}
\usepackage[prefixflatinterpret,prefixflatbracketmodalinterpret,prefixflatsetinterpret,sidenotecalculus,longseqcontext,silentconst]{logic}
\usepackage[prefixflatinterpret,prefixflatbracketmodalinterpret,differentialdL,simplenames,silentconst]{dL}

\usepackage{prettyref}
\newcommand{\rref}[2][]{\prettyref{#2}}
\newrefformat{sec}{Section\,\ref{#1}}
\newrefformat{subsec}{Section\,\ref{#1}}
\newrefformat{def}{Def.\,\ref{#1}}
\newrefformat{thm}{Theorem\,\ref{#1}}
\newrefformat{prop}{Proposition\,\ref{#1}}
\newrefformat{lem}{Lemma\,\ref{#1}}
\newrefformat{cor}{Corollary\,\ref{#1}}
\newrefformat{ex}{Example\,\ref{#1}}
\newrefformat{tab}{Table\,\ref{#1}}
\newrefformat{fig}{Fig.\,\ref{#1}}
\newrefformat{app}{Appendix\,\ref{#1}}
\newrefformat{alg}{Algorithm\,\ref{#1}}
\newrefformat{model}{Model\,\ref{#1}}
\newrefformat{line}{Line\,\ref{#1}}
\newrefformat{eq}{Eq.\,\eqref{#1}}
\newrefformat{case}{Case\,\eqref{#1}}
\newrefformat{appendix}{Appendix\,\ref{#1}}

\usepackage{float}
\floatstyle{ruled}
\newfloat{model}{tbp}{model}
\floatname{model}{Model}
\newcounter{modelline}
\makeatletter
\newcommand{\mline}[1]{{\refstepcounter{modelline}\ltx@label{#1}}~\text{\scriptsize{\themodelline}}\quad}
\makeatother

\makeatletter
\newcommand{\boxedSep}[2][\fboxsep]{{%
  \setlength{\fboxsep}{#1}\fbox{\m@th$\displaystyle#2$}}}
\makeatother

\newcommand{\highlightSynth}[1]{\boxedSep[0.3\fboxsep]{\,#1\,}}
\newcommand{\listingFrac}[2]{{#1}/{#2}}

\newif\iftacasversion
\tacasversionfalse

\makeatletter
\newcommand{\tacastext}[2]{%
  \iftacasversion
    #1
  \else
    #2
  \fi
}
\makeatother

\newcommand{\Sem}[1]{\llbracket #1 \rrbracket}

\newcommand{\afterProofE}{\medskip \noindent}

\newcommand{\UniformlyOptimal}[4]{{\textsf{best}}_{#1}(#2, #3, #4)} 

\newcommand{\tagged}[2]{{#1}^{(#2)}}

\DeclareMathOperator{\BV}{BV}

\newcommand{\probFilled}[1]{\prob(#1)}

\newcommand{\Subst}[2]{{#1}[#2]}
\newcommand{\SubstPair}[2]{{#1} \mapsto {#2}}

\newcommand{\Define}{\ \equiv \ }
\newcommand{\define}{\, \equiv \,}

\newcommand{\true}{\text{true}}
\newcommand{\false}{\text{false}}

\newcommand{\Goperator}[1]{\mathcal{G}(#1)}

\newcommand{\invHole}{\text{\Huge \textvisiblespace}}
\newcommand{\guardHole}[1]{\invHole_{\,#1}}
\newcommand{\drepeatn}[2]{{#1}^{\times \le #2}}
\newcommand{\prepeatn}[2]{{#1}^{* \le #2}}

\DeclareMathOperator{\Opt}{opt}
\newcommand{\Iopt}{I^{\hspace{0.1em}\Opt}}
\newcommand{\Uopt}{U^{\hspace{0.1em}\Opt}}
\newcommand{\Gopt}{G^{\hspace{0.1em}\Opt}}

\newcommand{\Sopt}{S^{\hspace{0.1em}\Opt}}

\newcommand{\Inth}[1]{I^{\hspace{0.05em} #1}}
\newcommand{\Gnth}[1]{G^{\hspace{0.05em} #1}}

\newcommand{\Reach}[2]{R(#1, #2)}
\newcommand{\ReachNoarg}{R}

\newcommand{\kwd}[1]{{\normalfont \textsf{#1}}\xspace} 
\newcommand{\otherdomain}{\kwd{domain}}
\newcommand{\prob}{\kwd{prob}}
\newcommand{\step}{\kwd{step}}
\newcommand{\forever}{\kwd{forever}}
\newcommand{\safe}{\kwd{safe}}
\newcommand{\assum}{\kwd{assum}}
\newcommand{\ctrlable}{\kwd{ctrlable}}
\newcommand{\plant}{\kwd{plant}}
\newcommand{\ctrl}{\kwd{ctrl}}
\newcommand{\act}{\kwd{act}}
\newcommand{\skp}{\kwd{skip}}

\newcommand{\perma}{\textsf{P}}

\newcommand{\reduce}{\kwd{reduce}}
\newcommand{\exact}{\kwd{exact}}

\newcommand{\parens}[1]{(#1)}
\newcommand{\curly}[1]{\{#1\}}

\newcommand{\InferEquiv}{\ \equiv \ }

\newcommand{\lequiv}{\leftrightarrow}

\renewcommand{\dbox}[2]{[#1] \, #2}  
\renewcommand{\ddiamond}[2]{\langle #1 \rangle \, #2}  

\newcommand{\seq}{\,;\,}
\newcommand{\D}[1]{\ensuremath{#1'}}

\newcommand*{\props}{\ensuremath{\mathcal{P}_{\hspace{0.1em}\reals}}\xspace}
\newcommand*{\fol}{\ensuremath{\text{FOL}_\reals}\xspace}

\newcommand{\dfree}[1]{{#1}^{-\mathsf{d}}}

\newcommand*{\fst}{\texttt{first}\xspace}

\newcommand{\abductp}{\ensuremath{\triangleright}\xspace}

\newcommand{\odereduce}{\ensuremath{\texttt{odereduce}}\xspace}

\def\leftrule{L}%
\def\rightrule{R}%

\newcommand{\I}{\dLint[state=\omega]}

\newcommand*{\synthesis}{Control Envelope Synthesis\xspace}

\newcommand{\anntnRemark}{The algebraic formulas presented are synthesized by CESAR automatically.
The annotations describing their meaning are added manually for the convenience of the reader.}

\begin{document}

\newcommand{\AlgoName}{CESAR\xspace}
\newcommand{\AlgoLong}{Control Envelope Synthesis via Angelic Refinements\xspace}

\title{CESAR: Control Envelope Synthesis via\\ Angelic Refinements \thanks{%
This work was funded by the Federal Railroad Administration Office of Research, Development and Technology under contract number 693JJ620C000025, a Swartz Center Innovation Commercialization Fellowship, and an Alexander von Humboldt Professorship.\\
This version of the contribution has been accepted for publication but is not the Version of Record and does not reflect post-acceptance improvements, or any corrections. The Version of Record is
available online at: http://dx.doi.org/10.1007/978-3-031-57246-3}.}

\author{
 Aditi Kabra\inst{1}\orcidID{0000-0002-2252-0539} \and
 Jonathan Laurent\inst{1,2}\orcidID{0000-0002-8477-1560} \and
 Stefan Mitsch\inst{1,3}\orcidID{0000-0002-3194-9759} \and
 Andr\'e Platzer\inst{1,2}\orcidID{0000-0001-7238-5710}}

\institute{Carnegie Mellon University, Pittsburgh, USA\\
\email{akabra@cs.cmu.edu}
\and Karlsruhe Institute of Technology, Karlsruhe, Germany\\
\email{$\{$jonathan.laurent,platzer$\}$@kit.edu}
\and DePaul University, Chicago, USA\\
\email{smitsch@depaul.edu}}
\maketitle

\begin{abstract}
This paper presents an approach for synthesizing provably correct control envelopes for hybrid systems.
Control envelopes characterize families of safe controllers and are used to monitor untrusted controllers at runtime.
Our algorithm fills in the blanks of a hybrid system's sketch specifying the desired shape of the control envelope, the possible control actions, and the system's differential equations.
In order to maximize the flexibility of the control envelope, the synthesized conditions saying which control action can be chosen when should be as permissive as possible while establishing a desired safety condition from the available assumptions, which are augmented if needed.
An implicit, optimal solution to this synthesis problem is characterized using hybrid systems game theory, from which explicit solutions can be derived via symbolic execution and sound, systematic game refinements.
Optimality can be recovered in the face of approximation via a dual game characterization.
The resulting algorithm, \emph{\AlgoLong (\AlgoName)}, is demonstrated in a range of safe control envelope synthesis examples with different control challenges.
\end{abstract}

\keywords{Hybrid systems \and Program synthesis \and Differential game logic}

\definecolor{vblue}{rgb}{.1,.15,.62}

\newsavebox{\Rval}%
\sbox{\Rval}{$\scriptstyle\mathbb{R}$}
\newsavebox{\backiterateb}%
\sbox{\backiterateb}{$\scriptstyle\overleftarrow{\dibox{{}^*}}$}

\irlabel{qear|\usebox{\Rval}}
\irlabel{qe|QE}
\irlabel{notr|$\lnot$\rightrule}
\irlabel{notl|$\lnot$\leftrule}
\irlabel{orr|$\lor$\rightrule}
\irlabel{orl|$\lor$\leftrule}
\irlabel{andr|$\land$\rightrule}
\irlabel{andl|$\land$\leftrule}
\irlabel{implyr|$\limply$\rightrule}
\irlabel{implyri|$\limply$\rightrule i}
\irlabel{implyl|$\limply$\leftrule}
\irlabel{leq|$\leq$}
\irlabel{equivr|$\lbisubjunct$\rightrule}
\irlabel{equivl|$\lbisubjunct$\leftrule}
\irlabel{id|id}
\irlabel{cut|cut}
\irlabel{assignrl|;}
\irlabel{cuprl|$\cup$}
\irlabel{refl|refl}
\irlabel{testrl|?}
\irlabel{cupl|$\cup_l$}
\irlabel{looprl|loop$_r$}
\irlabel{weakenr|W\rightrule}
\irlabel{weakenl|W\leftrule}
\irlabel{diamond|$\didia{\cdot}$}
\irlabel{box|$\dibox{\cdot}$}
\irlabel{composed|$\didia{{;}}$}
\irlabel{composeb|$\dibox{{;}}$}
\irlabel{choiced|$\didia{\cup}$}
\irlabel{choiceb|$\dibox{\cup}$}
\irlabel{iterated|$\didia{*}$}
\irlabel{iterateb|$\dibox{*}$}
\irlabel{testd|$\didia{?}$}
\irlabel{testb|$\dibox{?}$}
\irlabel{assignd|$\didia{:=}$}
\irlabel{assignb|$\dibox{:=}$}
\irlabel{evolved|$\didia{'}$}
\irlabel{evolveb|$\dibox{'}$}
\irlabel{Dassignb|$\dibox{':=}$}
\irlabel{duald|$\didia{{^d}}$}
\irlabel{dualb|$\dibox{{^d}}$}
\irlabel{dchoiced|$\didia{\cap}$}
\irlabel{dchoiceb|$\dibox{\cap}$}
\irlabel{K|K}
\irlabel{I|I}
\irlabel{DW|DW}
\irlabel{diffweaken|DW}
\irlabel{DI|DI}
\irlabel{DE|DE}
\irlabel{DC|DC}
\irlabel{DG|DG}
\irlabel{DS|DS}
\irlabel{Dconst|$c'$}
\irlabel{Dvar|$x'$}
\irlabel{Dplus|$+'$}
\irlabel{Dtimes|$\cdot'$}
\irlabel{Dcompose|$\compose'$}
\irlabel{loop|loop}
\irlabel{M|M}
\irlabel{Mb|M${\dibox{\cdot}}$}
\irlabel{Mbr|M\rightrule}
\irlabel{Mbl|M\leftrule}
\irlabel{V|V}
\irlabel{B|B}
\irlabel{Kd|K${\langle}{\rangle}$}
\irlabel{G|G}
\irlabel{Gr|G}
\irlabel{CT|CT}
\irlabel{CQ|CQ}
\irlabel{CE|CE}
\irlabel{CTr|CT\rightrule}
\irlabel{CTl|CT\leftrule}
\irlabel{CQr|CQ\rightrule}
\irlabel{CQl|CQ\leftrule}
\irlabel{CEr|CE\rightrule}
\irlabel{CEl|CE\leftrule}
\irlabel{invind|ind}
\irlabel{con|$con$}
\irlabel{congen|$con'$}
%
\irlabel{existsr|$\exists$\rightrule}
\irlabel{existsrinst|$\exists$\rightrule}
\irlabel{alll|$\forall$\leftrule}
\irlabel{alllinst|$\forall$\leftrule}
\irlabel{allr|$\forall$\rightrule}
\irlabel{existsl|$\exists$\leftrule}
\irlabel{iallr|i$\forall$}
\irlabel{iexistsr|i$\exists$}
\irlabel{ifb|$\dibox{\text{if}}$}
\irlabel{ifd|$\didia{\text{if}}$}
\irlabel{ifbl|$\dibox{\text{if}}$\leftrule}
\irlabel{ifbr|$\dibox{\text{if}}$\rightrule}
\irlabel{ifdl|$\didia{\text{if}}$\leftrule}
\irlabel{ifdr|$\didia{\text{if}}$\rightrule}
\irlabel{applyeqr|=\rightrule}
\irlabel{applyeql|=\leftrule}
\irlabel{dG|dG}
\irlabel{diffind|dI}
\irlabel{diffcut|dC}
\irlabel{dW|dW}
\irlabel{dA|dA}
\irlabel{difffin|dV}
\irlabel{band|${[]\land}$}
\irlabel{min|min}


\section{Introduction}
\label{sec:intro}
Hybrid systems are important models of many applications, capturing their differential equations and control \cite{LunzeLL09,Tabuada09,Alur15,Platzer18,DBLP:conf/eucc/AmesCENST19,Mitra21}.
For overall system safety, the correctness of the control decisions in a hybrid system is crucial.
Formal verification techniques can justify correctness properties.
Such correct controllers have been identified in a sequence of challenging case studies \cite{DBLP:conf/icfem/PlatzerQ09,squires18,DBLP:reference/mc/DoyenFPP18,pek2020fail,DBLP:conf/hybrid/IvanovCWAPL20,DBLP:conf/memocode/FreibergerSHA21,train}.
A useful approach to verified control is to design and verify a safe \textit{control envelope} around possible safe control actions.
Safe control envelopes are nondeterministic programs whose every execution is safe.
In contrast with controllers, control envelopes define entire families of controllers to allow control actions under as many circumstances as possible, as long as they maintain the safety of the hybrid system.
Safe control envelopes allow the verification of abstractions of control systems, isolating the parts relevant to the safety feature of interest, without involving the full complexity of a specific control implementation.
The full control system is then monitored for adherence to the safe control envelope at runtime~\cite{DBLP:journals/fmsd/MitschP16}.
The control envelope approach allows a single verification result to apply to multiple specialized control implementations, optimized for different objectives.
It puts industrial controllers that are too complex to verify directly within the reach of verification, because a control envelope only needs to model the safety-critical aspects of the controller.
Control envelopes also enable applications like justified speculative control \cite{saferl}, where machine-learning-based agents control safety-critical systems safeguarded within a verified control envelope, or \cite{DBLP:conf/nfm/QianM23}, where these envelopes generate reward signals for reinforcement learning.

Control envelope design is challenging.
Engineers are good at specifying the \emph{shape} of a model and listing the possible control actions by translating client specifications, which is crucial for the fidelity of the resulting model.
But identifying the exact control conditions required for safety in a model is a much harder problem that requires design insights and creativity, and is the main point of the deep area of control theory.
Most initial system designs are incorrect and need to be fixed before verification succeeds.
Fully rigorous justification of the safety of the control conditions requires full verification of the resulting controller in the hybrid systems model.
We present a synthesis technique that addresses this hard problem by filling in the holes of a hybrid systems model to identify a correct-by-construction control \emph{envelope} that is as permissive as possible.

Our approach is called \emph{\AlgoLong (\AlgoName)}.
The idea is to implicitly characterize the optimal safe control envelope via hybrid games yielding maximally permissive safe solutions in differential game logic \cite{Platzer18}.
To derive explicit solutions used for controller monitoring at runtime, we successively refine the games while preserving safety and, if possible, optimality.
Our experiments demonstrate that \AlgoName solves hybrid systems synthesis challenges requiring different control insights.

\paragraph{Contributions.}
The primary contributions of this paper behind \AlgoName are:
\begin{itemize}
\item optimal hybrid systems control envelope synthesis via hybrid games.
\item differential game logic formulas identifying optimal safe control envelopes.
\item refinement techniques for safe control envelope approximation, including \emph{bounded fixpoint unrollings} via a recurrence, which exploits \emph{action permanence} (a hybrid analogue to idempotence).
\item a primal/dual game counterpart optimality criterion.
\end{itemize}


\section{Background: Differential Game Logic}
\label{sec:background}

We use hybrid games written in differential game logic (\dGL, \cite{Platzer18}) to represent solutions to the synthesis problem.
Hybrid games are two-player noncooperative zero-sum sequential games with no draws that are played on a hybrid system with differential equations.
Players take turns and in their turn can choose to act arbitrarily within the game rules.
At the end of the game, one player wins, the other one loses.
The players are classically called Angel and Demon.
\textit{Hybrid systems}, in contrast, have no agents, only a nondeterministic controller running in a nondeterministic environment.
The synthesis problem consists of filling in holes in a hybrid system.
Thus, expressing solutions for hybrid \textit{system} synthesis with hybrid \textit{games} is one of the insights of this paper.

An example of a game is \({(\pumod{v}{1} \cap \pumod{v}{-1}) \seq \curly{x' = v}}\).
In this game, first Demon chooses between setting velocity $v$ to 1, or to -1.
Then, Angel evolves position $x$ as \(x'=v\) for a duration of her choice.
Differential game logic uses modalities to set win conditions for the players.
For example, in the formula \(\dbox{(\pumod{v}{1} \cap \pumod{v}{-1}) \seq \curly{x' = v}}{x \ne 0}\), Demon wins the game when $x\neq 0$ at the end of the game and Angel wins otherwise.
The overall formula represents the set of states from which Demon can win the game, which is \(x\neq 0\)
because when \(x<0\), Demon has the \emph{winning strategy} to pick \(\pumod{v}{-1}\), so no matter how long Angel evolves \(x'=v\), \(x\) remains negative.
Likewise, when \(x>0\), Demon can pick \(\pumod{v}{1}\).
However, when \(x=0\), Angel has a winning strategy: to evolve \(x'=v\) for zero time, so that \(x\) remains zero regardless of Demon's choice.

We summarize \dGL's program notation (\rref{tab:hybridgames}). See \cite{Platzer18} for full exposition.
\begin{table}[b!]
  \caption{Hybrid game operators for two-player hybrid systems}
  \centering
  \begin{tabularx}{\linewidth}{lX}
    \toprule
       \multicolumn{1}{l@{~~}}{\textbf{Game}}
    & \multicolumn{1}{c}{\textbf{Effect}}
    \\
    \midrule
    \(\pumod{x}{\theta}\) & assign value of term $\theta$ to variable $x$\\
    \(\ptest{\psi}\) & Angel passes challenge if formula $\psi$ holds in current state, else loses immediately\\
    \(\bigl(\D{x_1}=\theta_1,\dots,\) & Angel evolves $x_i$ along differential equation system $\D{x_i} = \theta_i$\\
    \(\phantom{\big(}\D{x_n}=\theta_n ~\&~ \psi\bigr)\) & for choice of duration $\geq0$, loses immediately when violating~$\psi$\\
    \(\alpha;\beta\) & sequential game, first play hybrid game~$\alpha$, then hybrid game~$\beta$  \\
    \(\pchoice{\alpha}{\beta}\) & Angel chooses to follow either hybrid game~$\alpha$ or $\beta$\\
    \(\prepeat{\alpha}\) & Angel repeats hybrid game~$\alpha$, choosing to stop or go after each $\alpha$\\
    \(\pdual{\alpha}\) & dual game switches player roles between Angel and Demon\\
    \midrule
    \(\dchoice{\alpha}{\beta}\) & demonic choice \(\pdual{(\pchoice{\pdual{\alpha}}{\pdual{\beta}})}\) gives choice between $\alpha$ and $\beta$ to Demon\\
    \(\drepeat{\alpha}\) & demonic repetition \(\pdual{(\prepeat{(\pdual{\alpha})})}\) gives control of repetition to Demon\\
    \bottomrule
  \end{tabularx}
  \label{tab:hybridgames}
\end{table}
Assignment \(\pumod{x}{\theta}\) instantly changes the value of variable \(x\) to the value of \(\theta\).
Challenge \(\ptest{\psi}\) continues the game if \(\psi\) is satisfied in the current state, otherwise Angel loses immediately.
In continuous evolution \(\D{x}=\theta ~\&~ \psi\) Angel follows the differential equation \(\D{x}=\theta\) for some duration of her choice, but loses immediately on violating \(\psi\) at any time.
Sequential game \(\alpha;\beta\) first plays \(\alpha\) and when it terminates without a player having lost, continues with \(\beta\).
Choice \(\pchoice{\alpha}{\beta}\) lets Angel choose whether to play \(\alpha\) or \(\beta\).
For repetition \(\prepeat{\alpha}\), Angel repeats \(\alpha\) some number of times, choosing to continue or terminate after each round.
The dual game \(\pdual{\alpha}\) switches the roles of players.
For example, in the game \(\pdual{\ptest{\psi}}\), Demon passes the challenge if the current state satisfies \(\psi\), and otherwise loses immediately.

In games restricted to the structures listed above but without \(\pdual{\alpha}\), all choices are resolved by Angel alone with no adversary, and hybrid games coincide with hybrid systems in differential dynamic logic (\dL) \cite{Platzer18}.
We will use this restriction to specify the synthesis \textit{question}, the sketch that specifies the shape and safety properties of control envelopes.
But to characterize the \emph{solution} that fills in the blanks of the control envelope sketch, we use games where both Angel and Demon play.
Notation we use includes demonic choice \(\dchoice{\alpha}{\beta}\), which lets Demon choose whether to run \(\alpha\) or \(\beta\).
Demonic repetition \(\drepeat{\alpha}\) lets Demon choose whether to repeat \(\alpha\) choosing whether to stop or go at the end of every run.
We define \(\prepeatn{\alpha}{n}\) and \(\drepeatn{\alpha}{n}\) for angelic and demonic repetitions respectively of at most $n$ times.

In order to express properties about hybrid games, differential game logic formulas refer to the existence of winning strategies for objectives of the games (e.g., a controller has a winning strategy to achieve collision avoidance despite an adversarial environment).
The set of \dGL~formulas is generated by the following grammar (where \m{{\sim}\in\{<,\leq,=,\geq,>\}} and $\theta_1,\theta_2$ are arithmetic expressions in~\m{+,-,\cdot,/} over the reals, \(x\) is a variable, \(\alpha\) is a hybrid game):
\[
\phi \coloneqq \theta_1 \sim \theta_2
\mid \lnot \phi
\mid \phi \land \psi
\mid \phi \lor \psi
\mid \phi \limply \psi
\mid \lforall{x}{\phi}
\mid \lexists{x}{\phi}
\mid \dbox{\alpha}{\phi}
\mid \ddiamond{\alpha}{\phi}
\]
Comparisons of arithmetic expressions, Boolean connectives, and quantifiers over the reals are as usual.
The modal formula \(\ddiamond{\alpha}{\phi}\) expresses that player Angel has a winning strategy to reach a state satisfying \(\phi\) in hybrid game \(\alpha\).
Modal formula \(\dbox{\alpha}{\phi}\) expresses the same for Demon.
The fragment without modalities is first-order real arithmetic.
Its fragment without quantifiers is called \emph{propositional arithmetic} \props.
Details on the semantics of \dGL \tacastext{can be found in \cite{Platzer18}}{\cite{Platzer18} is recalled in \rref{app:proofs}, but we provide examples to give the intuition.
\(\dbox{\pchoice{\alpha}{\beta}}{\phi}\) expresses that Demon has a winning strategy when Angel chooses between $\alpha$ and $\beta$ to achieve $\phi$, while \(\dbox{\dchoice{\alpha}{\beta}}{\phi}\) expresses that Demon has a winning strategy to achieve $\phi$ when Demon chooses whether to play game $\alpha$ or $\beta$.
Correspondingly, \(\ddiamond{\dchoice{\alpha}{\beta}}{\phi}\) expresses that Angel has a winning strategy to achieve $\phi$ when Demon has a choice between $\alpha$ and $\beta$}.
A formula $\phi$ is \emph{valid}, written \(\entails\phi\), iff it is true in every state $\omega$.
States are functions assigning a real number to each variable.
For instance, \(\phi \limply \dbox{\alpha}{\psi}\) is valid iff, from all initial states satisfying $\phi$, Demon has a winning strategy in game $\alpha$ to achieve $\psi$.
\tacastext{}{Our proofs are syntactic derivations in the \dGL proof calculus, summarized in \rref{app:proofs} with a standard first-order logic sequent calculus.
A \emph{sequent} \(\lsequent{\Gamma}{\Delta}\) with a finite list of antecedent formulas $\Gamma$ and succedent formulas $\Delta$ is short for \(\landfold_{\phi\in\Gamma}\phi \limply \lorfold_{\psi\in\Delta} \psi\).
}%

\paragraph{Control Safety Envelopes by Example.}
In order to separate safety critical aspects from other system goals during control design, we abstractly describe the safe choices of a controller with safe control envelopes that deliberately underspecify when and how to exactly execute certain actions.
They focus on describing in which regions it is safe to take actions.
For example, \rref{model:etcs} designs a train control envelope \cite{DBLP:conf/icfem/PlatzerQ09} that must stop by the train by the \textit{end of movement authority} \(e\) located somewhere ahead, as assigned by the train network scheduler.
Past \(e\), there may be obstacles or other trains.
The train's control choices are to accelerate or brake as it moves along the track.
The goal of \AlgoName is to synthesize the framed formulas in the model, that are initially blank.
\begin{model}[tb]
\setcounter{modelline}{0}
\caption{The train ETCS model (slightly modified from \cite{DBLP:conf/icfem/PlatzerQ09}). Framed formulas are initially blank and are automatically synthesized by our tool as indicated.}\label{model:etcs}
\begin{spreadlines}{0.3ex}
\vspace{-\baselineskip}
\begin{align*}
  \text{\assum} &\,\big|\mline{line:train-assums} A > 0 \land B > 0 \land T > 0 \land v \ge 0 \land \\
  \text{\ctrlable} &\,\big|\mline{line:train-init} \highlightSynth{e-p>\listingFrac{v^2}{2B}} \limply [\{ \\
  \text{\ctrl} &\left|
  \begin{aligned}
    &\mline{line:train-acc} \qquad ( \quad \parens{\ptest{\, \highlightSynth{e - p > vT + \listingFrac{AT^2}{2} + \listingFrac{(v + AT)^2}{2B} }} \seq \pumod{a}{A}} \\
    &\mline{line:train-brake} \qquad \cup \  (\ptest{\, \highlightSynth{\vphantom{A \cdot T^2/2B}\true}} \seq \pumod{a}{-B}) \quad ) \seq \\
  \end{aligned}
  \right.\\
  \text{\plant} &\,\big|\mline{line:train-plant} \qquad (\pumod{t}{0} \seq \{p'=v, v'=a, t'=1 \ \& \ t \le T \land v \ge 0\}) \\
  \text{\safe} &\,\big|\mline{line:train-safety} \}^\ast] (e-p>0)\\[-5ex]
\end{align*}
\end{spreadlines}
\end{model}

\rref{line:train-safety} describes the \emph{safety property} that is to be enforced at all times: the train driving at position $p$ with velocity $v$ must not go past position $e$.
\rref{line:train-assums} lists \emph{modeling assumptions}: the train is capable of both acceleration (\(A{>}0\)) and deceleration (\(B{>}0\)), the controller latency is positive (\(T{>}0\)) and the train cannot move backwards as a product of braking (this last fact is also reflected by having $v \ge 0$ as a domain constraint for the plant on \rref{line:train-plant}).
These assumptions are fundamentally about the physics of the problem being considered.
In contrast, \rref{line:train-init} features a \emph{controllability assumption} that can be derived from careful analysis.
Here, this synthesized assumption says that the train cannot start so close to \(e\) that it won't stop in time even if it starts braking immediately.
\rref{line:train-acc} and \rref{line:train-brake} describe a train controller with two actions: accelerating (\(\pumod{a}{A}\)) and braking (\(\pumod{a}{-B}\)).
Each action is guarded by a synthesized formula, called an \emph{action guard} that indicates when it is safe to use.
Angel has control over which action runs, and adversarially plays with the objective of violating safety conditions.
But Angel's options are limited to only safe ones because of the synthesized action guards, ensuring that Demon still wins and the overall formula is valid.
In this case, braking is always safe whereas acceleration can only be allowed when the distance to end position \(e\) is sufficiently large.
Finally, the plant on \rref{line:train-plant} uses differential equations to describe the train's kinematics.
A timer variable $t$ is used to ensure that no two consecutive runs of the controller are separated by more than time $T$.
Thus, this controller is \emph{time-triggered}.

\paragraph{Overview of \AlgoName.}
\AlgoName first identifies the optimal solution for the blank of \rref{line:train-init}.
Intuitively, this blank should identify a \emph{controllable invariant}, which denotes a set of states where a controller with choice between acceleration and braking has some strategy (to be enforced by the conditions of \rref{line:train-acc} and \rref{line:train-brake}) that guarantees safe control forever.
Such states can be characterized by the following \dGL formula where Demon, as a proxy for the controller, decides whether to accelerate or brake: \(\dbox{\parens{(\pumod{a}{A} \cap \pumod{a}{-B}) \seq \plant}^*}{\safe}\) where \plant and \safe are from \rref{model:etcs}.
When this formula is true, Demon, who decides when to brake to maintain the safety contract, has a winning strategy that the controller can mimic.
When it is false, Demon, a perfect player striving to maintain safety, has no winning strategy, so a controller has no guaranteed way to stay safe either.

This \dGL formula provides an \emph{implicit} characterization of the optimal controllable invariant from which we derive an explicit formula in \props to fill the blank with using symbolic execution.
Symbolic execution solves a game following the axioms of \dGL to produce an equivalent \props formula (\rref{sec:reduce}).
However, our \dGL formula contains a loop, for which symbolic execution will not terminate in finite time.
To reason about the loop, we \emph{refine} the game, modifying it so that it is easier to symbolically execute, but still at least as hard for Demon to win so that the controllable invariant that it generates remains sound.
In this example, the required game transformation first restricts Demon's options to braking.
Then, it eliminates the loop using the observation that the repeated hybrid iterations \((\pumod{a}{-B};\plant)^*\) behave the same as just following the continuous dynamics of braking for unbounded time.
It replaces the original game with \(\pumod{a}{-B} \seq \pumod{t}{0} \seq \{p'=v, v'=a \ \& \ \land v \ge 0\}\), which is loop-free and easily symbolically executed.
Symbolically executing this game to reach safety condition \safe yields controllable invariant \(e-p>\frac{v^2}{2B}\) to fill the blank of \rref{line:train-init}.

Intuitively, this refinement (formalized in \rref{sec:single-action}) captures situations where the controller stays safe forever by picking a single control action (braking).
It generates the optimal solution for this example because braking forever is the dominant strategy: given any state, if braking forever does not keep the train safe, then certainly no other strategy will.
However, there are other problems where the dominant control strategy requires the controller to strategically switch between actions, and this refinement misses some controllable invariant states.
So we introduce a new refinement: bounded game unrolling via a recurrence (\rref{sec:unrolling}).
A solution generated by unrolling $n$ times captures states where the controller can stay safe by switching control actions up to $n$ times.

Having synthesized the controllable invariant, \AlgoName fills the action guards (\rref{line:train-acc} and \rref{line:train-brake}).
An action should be permissible when running it for one iteration maintains the controllable invariant.
For example, acceleration is safe to execute exactly when \([\pumod{a}{A};\plant]e-p>\frac{v^2}{2B}\).
We symbolically execute this game to synthesize the formula that fills the guard of \rref{line:train-acc}.


\section{Approach}
\label{sec:approach}

This section formally introduces the \emph{\AlgoLong (\AlgoName)} approach for hybrid systems control envelope synthesis.

\subsection{Problem Definition}\label{sec:problem}

We frame the problem of \emph{control envelope synthesis} in terms of filling in holes $\invHole$ in a problem of the following shape:
\begin{equation}\label{eq:prob}
  \prob \Define \assum \land \invHole \limply \dbox{
    \big({
      (\cup_i \, (\ptest{\guardHole{i}} \seq \act_i)) \seq
      \plant
    }\big)^*
  }{\safe}.
\end{equation}
Here, the control envelope consists of a nondeterministic choice between a finite number of guarded actions.
Each action $\act_i$ is guarded by a condition $\guardHole{i}$ to be determined in a way that ensures safety within a controllable invariant \cite{controlled-inv-1,controlled-inv-2} $\invHole$ to be synthesized also. The plant is defined by the following template:
\begin{equation}
  \plant \Define \pumod{t}{0} \seq \curly{\pevolvein{x' = f(x), \, t'=1 \,}{\, \otherdomain \land t \le T}}.
\end{equation}
This ensures that the plant must yield to the controller after time $T$ at most, where $T$ is assumed to be positive and constant. In addition, we make the following assumptions:
\begin{enumerate}
\item Components $\assum$, $\safe$ and $\otherdomain$ are propositional arithmetic formulas.
\item Timer variable $t$ is fresh (does not occur except where shown in template).
\item Programs $\act_i$ are discrete $\dL$ programs that can involve choices, assignments and tests with propositional arithmetic. Variables assigned by $\act_i$ must not appear in $\safe$. In addition, $\act_i$ must terminate in the sense that \(\entails \ddiamond{\act_i}{\true}\).
\item The modeling assumptions $\assum$ are invariant in the sense that $\entails \assum \limply \dbox{(\cup_i \, \act_i) \seq \plant}{\assum}$. This holds trivially for assumptions about constant parameters such as $A>0$ in \rref{model:etcs} and this ensures that the controller can always rely on them being true.
\end{enumerate}

\begin{definition} \label{def:synthesis-problem}
  A \emph{solution} to the synthesis problem above is defined as a pair $(I, G)$ where $I$ is a formula and $G$ maps each action index $i$ to a formula $G_i$. In addition, the following conditions must hold:
\begin{enumerate}
  \item Safety is guaranteed: $\probFilled{I, G} \equiv \Subst{\prob}{\SubstPair{\invHole}{I}, \SubstPair{\guardHole{i}}{G_i}}$ is valid and $(\assum \land I)$ is a loop invariant that proves it so.
  \item \label{case:some-act}
  There is always some action: $(\assum \land I) \limply \bigvee_i G_i$ is valid.
\end{enumerate}
\end{definition}
Condition~\ref{case:some-act} is crucial for using the resulting nondeterministic control envelope, since it guarantees that safe actions are always available as a fallback.

\subsection{An Optimal Solution}

Solutions to a synthesis problem may differ in quality.
Intuitively, a solution is better than another if it allows for a strictly larger controllable invariant. In case of equality, the solution with the more permissive control envelope wins.
Formally, given two solutions $S = (I, G)$ and $S' = (I', G')$, we say that $S'$ is better or equal to $S$ (written $S \sqsubseteq S'$) if and only if $\entails \assum \limply (I \limply I')$ and additionally either $\entails \assum \limply \neg (I' \limply I)$ or $\entails (\assum \land I) \limply \landfold_i \, (G_i \limply G'_i)$.
Given two solutions $S$ and $S'$, one can define a solution $S \sqcap S' = (I \lor I',\, i \mapsto (I \land G_i \,\lor\, I' \land G_i'))$ that is better or equal to both $S$ and $S'$ ($S \sqsubseteq S \sqcap S'$ and $S' \sqsubseteq S \sqcap S'$).
A solution $S'$ is called the \emph{optimal solution} when it is the maximum element in the ordering, so that for any other solution $S$, $S \sqsubseteq S'$.
The optimal solution exists and is expressible in $\dGL$:
\begin{align}
  \Iopt &\Define \dbox{\parens{(\cap_i \, \act_i) \seq \plant}^*}{\safe} \label{eq:def-istar} \\
  \Gopt_i &\Define \dbox{\act_i \seq \plant}{\Iopt}. \label{eq:def-gstar}
\end{align}
Intuitively, $\Iopt$ characterizes the set of all states from which an optimal controller (played here by Demon) can keep the system safe forever. In turn, $\Gopt$ is defined to allow any control action that is guaranteed to keep the system within $\Iopt$ until the next control cycle as characterized by a modal formula.
\rref{sec:controllable} formally establishes the correctness and optimality of $\Sopt \equiv (\Iopt, \, \Gopt)$.

While it is theoretically reassuring that an optimal solution exists that is at least as good as all others and that this optimum can be characterized in \dGL, such a solution is of limited practical usefulness since \rref{eq:def-istar} cannot be executed without solving a game at runtime.
Rather, we are interested in \emph{explicit} solutions where $I$ and $G$ are quantifier-free real arithmetic formulas. There is no guarantee in general that such solutions exist that are also optimal, but our goal is to devise an algorithm to find them in the many cases where they exist or find safe approximations otherwise.


\subsection{Controllable Invariants}\label{sec:controllable}

The fact that $\Sopt$ is a solution can be characterized in logic with the notion of a controllable invariant that, at each of its points, admits some control action that keeps the plant in the invariant for one round.
All lemmas and theorems throughout this paper are proved in \tacastext{the extended preprint \cite[Appendix B]{arxiv}}{\rref{app:proofs}}.

\begin{definition}[Controllable Invariant]
A \emph{controllable invariant} is a formula $I$ such that $\,\entails I \limply \safe$ and $\, \entails I \limply \lorfold_i\, \dbox{\act_i \seq \plant}{I}$.
\end{definition}

\noindent From this perspective, $\Iopt$ can be seen as the largest controllable invariant.

\begin{lemmaE}\label{lem:optimal-is-ctrlinv}
  $\Iopt$ is a controllable invariant and it is optimal in the sense that $\entails I \limply \Iopt$ for any controllable invariant $I$.
\end{lemmaE}
\begin{proofE}
  Let us first prove that $\Iopt \equiv \dbox{\parens{(\cap_i \, \act_i) \seq \plant}^*}{\safe}$ is a controllable invariant.
  First, note that axiom \irref{iterateb} along with the definition of $\Iopt$ derives
  \begin{equation}
  \infers \Iopt \lbisubjunct \safe \land \dbox{(\cap_i \, \act_i) \seq \plant}{\Iopt}
    \label{eq:opt-is-fixpoint}
  \end{equation}
  because that is
  \(\infers \Iopt \lbisubjunct \safe \land \dbox{(\cap_i \, \act_i) \seq \plant}{\dbox{\parens{(\cap_i \, \act_i) \seq \plant}^*}{\safe}}\).

Safety \(\infers \Iopt \limply \safe\) derives from \rref{eq:opt-is-fixpoint} propositionally.

Controllable invariance \(\infers \Iopt \limply \lorfold_i\, \dbox{\act_i \seq \plant}{\Iopt}\) derives from \rref{eq:opt-is-fixpoint}:
\begin{sequentdeduction}[array]
  \linfer
  {\linfer[weakenl+composeb]
    {\linfer[dchoiceb]
      {\linfer[id]
        {\lclose}
        {\lsequent{\lorfold_i\,\dbox{\act_i}{\dbox{\plant}{\Iopt}}}{\lorfold_i\, \dbox{\act_i}{\dbox{\plant}{\Iopt}}}}
      }
      {\lsequent{\dbox{\cap_i \, \act_i}{\dbox{\plant}{\Iopt}}}{\lorfold_i\, \dbox{\act_i}{\dbox{\plant}{\Iopt}}}}
    }
    {\lsequent{\safe \land \dbox{(\cap_i \, \act_i) \seq \plant}{\Iopt}}{\lorfold_i\, \dbox{\act_i \seq \plant}{\Iopt}}}
  }
  {\lsequent{\Iopt}{\lorfold_i\, \dbox{\act_i \seq \plant}{\Iopt}}}
\end{sequentdeduction}
This concludes the proof that $\Iopt$ is a controllable invariant.  Let us now prove that $\Iopt$ is optimal. That is, let us consider a controllable invariant $I$ and derive $\lsequent{}{I \limply \Iopt}$:
\begin{sequentdeduction}[array]
  \linfer[]
  {
    \linfer[loop]
    {
      \linfer[id]{\lclose}{\lsequent{I}{I}}!
      \linfer[composeb+dchoiceb+composeb]
      {
        \lsequent{I}{\lorfold_i \dbox{\act_i \seq \plant}{I}}
      }
      {
        \lsequent{I}{\dbox{(\cap_i \, \act_i) \seq \plant}{I}}
      }!
      \lsequent{I}{\safe}
    }
    {
      \lsequent{I}{\dbox{\parens{(\cap_i \, \act_i) \seq \plant}^*}{\safe}}
    }
  }
  {
    \lsequent{}{I \limply \Iopt}
  }
\end{sequentdeduction}
The two remaining premises are the two parts of the definition of $I$ being a controllable invariant. This concludes the proof.\qed
\end{proofE}

\afterProofE Moreover, not just $\Iopt$, but \emph{every} controllable invariant induces a solution.
Indeed, given a controllable invariant $I$, we can define
\(\Goperator{I} \equiv (i \mapsto \dbox{\act_i \seq \plant}{I})\) for the \emph{control guards induced by} $I$.
$\Goperator{I}$ chooses as the guard for each action $\act_i$ the modal condition ensuring that $\act_i$, preserves $I$ after the $\plant$.

\begin{lemmaE}\label{lem:controllable-inv-sol}
  If $I$ is a controllable invariant, then $(I, \Goperator{I})$ is a solution (\rref{def:synthesis-problem}).
\end{lemmaE}
\begin{proofE}
Let us assume that $I$ is a controllable invariant and prove that $(I, G)$ is a solution with $G \equiv \Goperator{I}$.
We first need to prove that $\assum \land I$ is an invariant for $\probFilled{I, G}$.
Since $\assum$ is already assumed to be an invariant, it is enough to prove that $I$ is an invariant itself.
$I$ holds initially since $I \limply I$ is valid and it implies $\safe$ by definition of a controllable invariant. Preservation holds by the definition of $G$:
\begin{sequentdeduction}[array]
  \linfer[composeb+choiceb+testb+andr]{
    \linfer[]{
      \linfer[id]{\lclose{}}{\lsequent{I, G_i}{G_i}}
    }{
      \lsequent{I, G_i}{\dbox{\act_i \seq \plant}{I}}
    }
  }{
    \lsequent{}{I \limply \dbox{\cup_i \, (\ptest{G_i} \seq \act_i) \seq \plant}{I}}
  }
\end{sequentdeduction}
where the axioms \irref{composeb+choiceb+testb} are used to unpack and repack the games leading to one conjunct for each action (treated separately via \irref{andr}).
We also need to prove that an action is always available, which holds by virtue of $I$ being a controllable invariant:
\begin{sequentdeduction}[array]
  \linfer[]{
    \lsequent{}{I \limply \lorfold_i \, \dbox{\act_i \seq \plant}{I}}
  }{
  \lsequent{}{I \limply \lorfold_i \, G_i}}
\end{sequentdeduction}
\qed
\end{proofE}

\afterProofE Conversely, a controllable invariant can be derived from any solution.

\begin{lemmaE}\label{lem:solution-implies-better-controllable}
  If $(I, G)$ is a solution, then $I' \equiv (\assum \land I)$ is a controllable invariant. Moreover, we have $(I, G) \sqsubseteq (I', \Goperator{I'})$.
\end{lemmaE}
\begin{proofE}
  Consider a solution $(I, G)$.
  We prove that $I' \equiv (\assum \land I)$ is a controllable invariant.
  Per \rref{def:synthesis-problem}, $I'$ is an invariant for $\probFilled{I, G}$ and so $\entails I' \limply \safe$.
  Also, we have the following derivation which repacks games via axioms \irref{choiceb+testb+composeb} using their equivalences:

\begin{sequentdeduction}[array]
  \linfer[cut]
  {
    \lsequent{I'}{\lorfold_i G_i}
    !
    \linfer[orl+M]
    {
      \linfer[implyr+testb+composeb] {
        \linfer[M]
        {
          \linfer[choiceb]{
            \lsequent{I'}{\dbox{\cup_i (\ptest{G_i} \seq \act_i) \seq \plant}{I'}}
          }{
            \lsequent{I'}{\landfold_i \, \dbox{\ptest{G_i} \seq \act_i \seq \plant}{I'}}
          }
        }
        {\lsequent{I'}{\dbox{\ptest{G_i} \seq \act_i \seq \plant}{I'}}}
      }
      {
        \lsequent{I', G_i}{\dbox{\act_i \seq \plant}{I'}}
      }
    }
    {
      \lsequent{I', \lorfold_i G_i}{\lorfold_i \dbox{\act_i \seq \plant}{I'}}
    }
  }
  {
    \lsequent{I'}{\lorfold_i \dbox{\act_i \seq \plant}{I'}}
  }
\end{sequentdeduction}
where the open premises are part of the definition of $(I, G)$ being a solution according to \rref{def:synthesis-problem}.
Let us now prove that $(I, G) \sqsubseteq (I', \Goperator{I'})$.
Trivially, we have $\entails \assum \limply (I \limply (\assum \land I))$.
Let us now derive $\entails \assum \land I \limply \landfold_i (G_i \limply \Goperator{I'}_i)$:
\begin{sequentdeduction}[array]
  \linfer[]
  {
    \linfer[choiceb+testb+composeb]
    {
        \lsequent{\assum \land I}{\dbox{(\cup_i \, (\ptest{G_i} \seq \act_i)) \seq \plant}{(\assum \land I)}}
    }
    {
      \lsequent{\assum \land I}{\landfold_i (G_i \limply \dbox{\act_i \seq \plant}{(\assum \land I)})}
    }
  }
  {
    \lsequent{\assum \land I}{\landfold_i (G_i \limply \Goperator{I'}_i)}
  }
\end{sequentdeduction}
where the remaining premise is part of the definition of $(I, G)$ being a solution. This concludes the proof. \qed

\end{proofE}

\afterProofE Solution comparisons w.r.t.\ $\sqsubseteq$ reduce to implications for controllable invariants.

\begin{lemmaE}\label{lem:compare-controllable}
  If $I$ and $I'$ are controllable invariants, then \((I, \Goperator{I}) \sqsubseteq (I', \Goperator{I'})\) if and only if \(\,\entails \assum \limply (I \limply I')\).
\end{lemmaE}
\begin{proofE}
  Let us first assume that $\assum \entails I \limply I'$ and prove that $(I, \Goperator{I}) \sqsubseteq (I', \Goperator{I'})$.
  It remains to show either $\assum \entails \neg(I' \limply I)$ or $(\assum \land I) \limply \landfold_i \, (G_i \limply G'_i)$ is valid.
  We show that $\assum \entails I \limply (\Goperator{I}_i \limply \Goperator{I'}_i)$ for all $i$.
  To do so, we leverage the fact that $\assum$ is an invariant.
  \begin{sequentdeduction}[array]
    \linfer[]{
      \linfer[iterateb]{
        \linfer[band+M]{
          \linfer[]{\lclose}{\lsequent{\assum, \ I }{I'}}
        }{\lsequent{\dbox{\act_i \seq \plant}{\assum}, \ \dbox{\act_i \seq \plant}{I} }{\dbox{\act_i \seq \plant}{I'}}}
      }{
      \lsequent{\assum, \ \dbox{\act_i \seq \plant}{I} }{\dbox{\act_i \seq \plant}{I'}}}
    }{
    \lsequent{\assum}{I \limply (\Goperator{I}_i \limply \Goperator{I'}_i)}}
  \end{sequentdeduction}
  The reverse direction follows trivially from the definition of $\sqsubseteq$. \qed
\end{proofE}

\afterProofE Taken together, these lemmas allow us to establish the optimality of $\Sopt$.

\begin{theoremE}[][restate, text proof={}]
  \label{thm:star-optimal}
  $\Sopt$ is an optimal solution (i.e. a maximum w.r.t.\ $\sqsubseteq$) of \rref{def:synthesis-problem}.
\end{theoremE}
\begin{proofE}
  We have $\Sopt \equiv (\Iopt, \,\Goperator{\Iopt})$. From \rref{lem:optimal-is-ctrlinv} and \rref{lem:controllable-inv-sol}, $\Sopt$ is a solution. Let $(I, G)$ be another solution. From \rref{lem:solution-implies-better-controllable}, there exists a controllable invariant $I'$ such that $(I, G) \sqsubseteq (I', \Goperator{I'})$. Then, from \rref{lem:compare-controllable} and from the optimality of $\Iopt$ (\rref{lem:optimal-is-ctrlinv}), we have $(I', \Goperator{I'}) \sqsubseteq (\Iopt, \Goperator{\Iopt})$. By transitivity, $(I, G) \sqsubseteq \Sopt$. This concludes the proof.
  \qed
\end{proofE}

\afterProofE
This shows the roadmap for the rest of the paper:
finding solutions to the control envelope synthesis problem reduces to finding controllable invariants that imply $\Iopt$, which can be found by
restricting the actions available to Demon in $\Iopt$ to guarantee safety, thereby \emph{refining} the associated game.


\subsection{One-Shot Fallback Refinement}
\label{sec:single-action}

The simplest refinement of $\Iopt$ is obtained when fixing a single fallback action to use in all states (if that is safe).
A more general refinement considers different fallback actions in different states, but still only plays one such action forever.

Using the $\dGL$ axioms, any loop-free $\dGL$ formula whose ODEs admit solutions expressible in real arithmetic can be automatically reduced to an equivalent first-order arithmetic formula (in \fol). An equivalent propositional arithmetic formula  in \props can be computed via quantifier elimination (QE). For example:
\begin{align*}
  & \dbox{(\pumod{v}{1} \cap \pumod{v}{-1}) \seq \curly{x' = v}}{x \ne 0} \\
  \InferEquiv & \dbox{\pumod{v}{1} \cap \pumod{v}{-1}}{\dbox{\curly{x' = v}}{x \ne 0}} && \text{by \irref{composeb}} \\
  \InferEquiv \ & \dbox{\pumod{v}{1}}{\dbox{\curly{x' = v}}{x \ne 0}} \ \lor\  \dbox{\pumod{v}{-1}} \dbox{\curly{x' = v}}{x \ne 0} && \text{by \irref{dchoiceb}} \\
  \InferEquiv \ & \dbox{\curly{x' = 1}}{x \ne 0} \ \lor\  \dbox{\curly{x' = -1}} {x \ne 0} && \text{by \irref{assignb}} \\
  \InferEquiv \ &  (\forall t{\ge}0\, x + t \ne 0) \lor  (\forall t{\ge}0\, x - t \ne 0) && \text{by \irref{evolveb},\irref{assignb}} \\
  \InferEquiv \ & x > 0 \ \lor \  x < 0 && \text{by QE} \enspace .
\end{align*}
Even when a formula features nonsolvable ODEs, techniques exist to compute weakest preconditions for differential equations, with conservative approximations~\cite{DBLP:journals/fmsd/SogokonMTCP22} or even exactly in some cases~\cite{platzer2020differential,DBLP:conf/sofsem/Boreale18}. In the rest of this section and for most of this paper, we are therefore going to assume the existence of a \reduce oracle that takes as an input a loop-free \dGL formula and returns a quantifier-free arithmetic formula that is equivalent modulo some assumptions.
\rref{sec:reduce} shows how to implement and optimize \reduce.

\begin{definition}[Reduction Oracle]\label{def:reduce}
A \emph{reduction oracle} is a function $\reduce$ that takes as an input a loop-free \dGL formula $F$ and an assumption $A \in \props$. It returns a formula $R \in \props$ along with a boolean flag $\exact$ such that the formula \(A \limply (R \limply F)\) is valid,
and if $\exact$ is true, then \(A \limply (R \lequiv F)\) is valid as well.
\end{definition}

Back to our original problem, $\Iopt$ is not directly reducible since it involves a loop. However, conservative approximations can be computed by restricting the set of strategies that the Demon player is allowed to use. One extreme case allows Demon to only use a single action $\act_i$ repeatedly as a fallback (e.g. braking in the train example). In this case, we get a controllable invariant $\dbox{\prepeat{(\act_i \seq \plant)}}{\safe}$, which further simplifies into $\dbox{\act_i \seq \plant_\infty}{\safe}$ with
\[\plant_\infty \!\equiv \curly{\pevolvein{ x' = f(x), t'=1\,}{\,\otherdomain}}\] a variant of \plant that never yields control. For this last step to be valid though, a technical assumption is needed on $\act_i$, which we call \emph{action permanence}.

\begin{definition}[Action Permanence]
An action $\act_i$ is said to be \emph{permanent} if and only if $(\act_i \seq \plant \seq \act_i) \equiv (\act_i \seq \plant)$, i.e., they are equivalent games.
\end{definition}

Intuitively, an action is \emph{permanent} if executing it more than once in a row has no consequence for the system dynamics. This is true in the common case of actions that only assign constant values to control variables that are read but not modified by the plant, such as $\pumod{a}{A}$ and $\pumod{a}{-B}$ in \rref{model:etcs}.

\begin{lemmaE}
  If $\act_i$ is permanent, 
  $ \entails \dbox{\prepeat{(\act_i \seq \plant)}}{\safe} \lequiv \dbox{\act_i \seq \plant_\infty}{\safe}$.
\end{lemmaE}
\begin{proofE}
We first prove that $ (\act_i \seq \plant)^n \equiv (\act_i \seq \plant^n)$ by induction on $n \ge 1$. The base case is trivial. Regarding the induction case, we have
\begin{align*}
  (\act_i \seq \plant)^{n+1}
  \InferEquiv & (\act_i \seq \plant \seq (\act_i \seq \plant)^{n}) \\
  \InferEquiv & (\act_i \seq \plant \seq \act_i \seq \plant^{n}) \\
  \InferEquiv & (\act_i \seq \plant \seq \act_i \seq \plant^{n}) \\
  \InferEquiv & (\act_i \seq \plant \seq \plant^{n}) \\
  \InferEquiv & (\act_i \seq \plant^{n+1}).
\end{align*}
From this, we get $(\act_i \seq \plant)^* \define \ptest{\true} \cup (\act_i \seq \plant^*)$ from the semantics of loops in \dGL. Thus, we have $ \entails \dbox{\prepeat{(\act_i \seq \plant)}}{\safe} \ \lequiv \  \safe \land \dbox{\act_i \seq \plant_\infty}{\safe}$ since $t$ does not appear free in $\safe$. From this, we prove our theorem by noting that $\safe \land \dbox{\act_i \seq \plant_\infty}{\safe} \ \lequiv \  \dbox{\act_i \seq \plant_\infty}{\safe} $ since $\act_i$ cannot write any variable that appears in $\safe$. \qed
\end{proofE}

\afterProofE Our discussion so far identifies the following approximation to our original synthesis problem, where $\perma$ denotes the set of all indexes of permanent actions:
\begin{align*}
  \Inth{0} &\Define \dbox{(\cap_{i\in\perma} \, \act_i) \seq \plant_\infty }{\safe}, \\
  \Gnth{0}_i &\Define \dbox{\act_i \seq \plant}{\Inth{0}}.
\end{align*}
Here, $\Inth{0}$ encompasses all states from which the agent can guarantee safety indefinitely with a single permanent action. $\Gnth{0}$ is constructed according to $\Goperator{\Inth{0}}$ and only allows actions that are guaranteed to keep the agent within $\Inth{0}$ until the next control cycle. Note that $\Inth{0}$ degenerates to $\false$ in cases where there are no permanent actions, which does not make it less of a controllable invariant.

\begin{theoremE}[][restate]
  \label{thm:s0-solution}
  $\Inth{0}$ is a controllable invariant.
\end{theoremE}

\begin{proofE}
  Trivially, we have $\entails \Inth{0} \limply \safe$. More interestingly, let us prove that $\Inth{0} \limply \lor_i \, \dbox{\alpha_i}{\Inth{0}}$ where $\alpha_i \define (\act_i \seq \plant)$. The proof crucially leverages the permanence assumption via the identity $\entails \Inth{0} \lequiv \lor_{i \in \perma} \, \dbox{\prepeat{\alpha_i}}{\safe}$.
\begin{sequentdeduction}[array]
  \linfer[weakenr]{
    \linfer[]{
      \linfer[orl+Mb]{
        \linfer[]{
          \linfer[Mb]{
            \linfer[iterateb]{\lclose}
            {\lsequent{\dbox{\prepeat{\alpha_i}}{\safe}}{\dbox{\alpha_i}{\dbox{\prepeat{\alpha_i}}{\safe}}}}}
          {\lsequent{\dbox{\prepeat{\alpha_i}}{\safe}}{\dbox{\alpha_i}{(\lor_{j \in \perma} \,\dbox{\prepeat{\alpha_j}}{\safe})}}}}
        {\lsequent{\dbox{\prepeat{\alpha_i}}{\safe}}{\dbox{\alpha_i}{\Inth{0}}}}}
      {\lsequent{\lor_{i \in \perma} \, \dbox{\prepeat{\alpha_i}}{\safe}}{\lor_{i \in \perma} \, \dbox{\alpha_i}{\Inth{0}}}}
    }{\lsequent{\Inth{0}}{\lor_{i \in \perma} \, \dbox{\alpha_i}{\Inth{0}}}}
  }{\lsequent{\Inth{0}}{\lor_i \, \dbox{\alpha_i}{\Inth{0}}}}
\end{sequentdeduction} \qed
\end{proofE}

\afterProofE Moreover, in many examples of interest, $\Inth{0}$ and $\Iopt$ are equivalent since an optimal fallback strategy exists that only involves executing a single action. This is the case in particular for \rref{model:etcs}, where
\begin{align*}
  \Inth{0} & \InferEquiv \dbox{\pumod{a}{-B} \seq \{p'=v, v'=a \ \& \  v \ge 0\}}{e - p > 0} \\
  & \InferEquiv \, e - p > v^2/2B
\end{align*}
characterizes all states at safe braking distance to the obstacle and $\Gnth{0}$ associates the following guard to the acceleration action:
\begin{align*}
  \Gnth{0}_{\pumod{a}{A}} & \InferEquiv \dbox{\pumod{a}{A} \seq \{p'=v, v'=a, t'=1 \ \& \  v \ge 0 \land t \le T\}}{e - p > v^2/2B} \\
  & \InferEquiv \, e - p > vT + \listingFrac{AT^2}{2} + \listingFrac{(v + AT)^2}{2B}
\end{align*}
That is, accelerating is allowed if doing so is guaranteed to maintain sufficient braking distance until the next control opportunity.
\rref{sec:optimality} discusses automatic generation of a proof that $(\Inth{0}, \Gnth{0})$ is an optimal solution for \rref{model:etcs}.


\subsection{Bounded Fallback Unrolling Refinement}
\label{sec:unrolling}

In \rref{sec:single-action}, we derived a solution by computing an underapproximation of $\Iopt$ where the fallback controller (played by Demon) is only allowed to use a one-shot strategy that picks a single action and plays it forever. Although this approximation is always safe and, in many cases of interest, happens to be exact, it does lead to a suboptimal solution in others. In this section, we allow the fallback controller to switch actions a bounded number of times before it plays one forever.
There are still cases where doing so is suboptimal (imagine a car on a circular race track that is forced to maintain constant velocity). But this restriction is in line with the typical understanding of a fallback controller, whose mission is not to take over a system indefinitely but rather to maneuver it into a state where it can safely get to a full stop~\cite{pek2020fail}.

For all bounds $n \in \mathbb{N}$, we define a game where the fallback controller (played by Demon) takes at most $n$ turns to reach the region $I^0$ in which safety is guaranteed indefinitely. During each turn, it picks a permanent action and chooses a time $\theta$ in advance for when it wishes to play its next move. Because the environment (played by Angel) has control over the duration of each control cycle, the fallback controller cannot expect to be woken up after time $\theta$ exactly. However, it can expect to be provided with an opportunity for its next move within the $[\theta, \theta+T]$ time window since the plant can never execute for time greater than $T$. Formally, we define $\Inth{n}$ as follows:
\begin{align*}
  \Inth{n} &\,\equiv\, \dbox{\drepeatn{\step}{n} \seq \forever}{\safe} \qquad \forever \,\equiv\, (\cap_{i\in\perma} \, \act_i) \seq \plant_\infty \\
  \step &\,\equiv\, \pdual{(\prandom{\theta} \seq \ptest{\theta \ge 0})} \seq (\cap_{i\in\perma} \, \act_i) \seq \plant_{\theta + T} \seq \dtest{\safe} \seq \ptest{t\ge\theta}
\end{align*}
where $\plant_{\theta+T}$ is the same as $\plant$, except that the domain constraint $t\le T$ is replaced by $t\le \theta+T$. Equivalently, we can define $I^n$ by induction as follows:
\begin{equation}
  \Inth{n+1} \define \Inth{n} \,\lor\, \dbox{\step}{\Inth{n}} \qquad \Inth{0} \define \dbox{\forever}{\safe},
\end{equation}
where the base case coincides with the definition of $\Inth{0}$ in \rref{sec:single-action}. Importantly, $\Inth{n}$ is a loop-free controllable invariant and so $\reduce$ can compute an explicit solution to the synthesis problem from $\Inth{n}$.

\begin{theoremE}
  \label{thm:sn-solution}
$\Inth{n}$ is a controllable invariant for all $n \ge 0$.
\end{theoremE}
\begin{proofE}
  We proceed by induction on $n$. The base case is covered by Theorem~\ref{thm:s0-solution}. Assume that $\Inth{n}$ is a controllable invariant and prove that $\Inth{n+1}$ is one also.
  Abbreviate $\alpha_i \define (\act_i \seq \plant)$.  Without loss of generality, assume that all actions are permanent since non-permanent actions play no role in computing $\Inth{n}$.  The hard part is in proving that $\entails \Inth{n} \limply \lor_i \, \dbox{\alpha_i}{\Inth{n+1}}$.
\begin{sequentdeduction}[array]
  \linfer[]{
    \linfer[orl]{
      \linfer[Mb]{
        \lsequent{\Inth{n}}{\lor_i \, \dbox{\alpha_i}{I^{n}}}!
        \lsequent{}{\Inth{n} \limply \Inth{n+1}}
      }{\lsequent{\Inth{n}}{\lor_i \, \dbox{\alpha_i}{\Inth{n+1}}}}!
      \lsequent{\dbox{\step}{\Inth{n}}}{\lor_i \, \dbox{\alpha_i}{\Inth{n+1}}}
    }{
      \lsequent{\Inth{n} \,\lor\, \dbox{\step}{\Inth{n}}}{\lor_i \, \dbox{\alpha_i}{\Inth{n+1}}}}}{
    \lsequent{\Inth{n+1}}{\lor_i \, \dbox{\alpha_i}{\Inth{n+1}}}}
\end{sequentdeduction}
The first premise is a consequence of $(\Inth{n}, \Gnth{n})$ being a solution (our induction hypothesis) and the second one is a trivial consequence of the definition of $\Inth{n+1}$. We can now focus on proving the last premise.

To do so, it is useful to introduce the following predicate:
\[\Reach{a}{b} \Define \dbox{\plant_{b}}{(\safe \land (t \ge a \limply \Inth{n}))}\]
Intuitively, $\Reach{a}{b}$ is true if following the dynamics leads to reaching $\Inth{n}$ within time interval $[a, b]$ while being safe the whole time. Using this predicate, we can reformulate $\dbox{\step}{\Inth{n}}$ as follows:
\begin{align}
  \label{eq:step-in-reach}
  \dbox{\step}{\Inth{n}} \,\equiv\, \lor_i \, \exists \theta\!\ge\!0 \, \dbox{\act_i}\Reach{\theta}{\theta+T}.
\end{align}
In addition, Lemma~\ref{lem:ode-reachability} gives us the following key property of $\ReachNoarg$:
\begin{align}
  \label{eq:reach-lem-app}
  \lsequent{}{c \le b \land \Reach{a}{b} \limply \dbox{\plant_c}{\Reach{a-t}{b-t}}}.
\end{align}
We can now complete the proof using~\rref{eq:step-in-reach}:
\begin{sequentdeduction}[array]
\linfer[]{
  \linfer[orr+orl]{
    \linfer[]{
      \lsequent{\Gamma}{\dbox{\alpha_i}{(t \le \theta \limply \Inth{n+1})}}!
      \lsequent{\Gamma}{\dbox{\alpha_i}{(t \ge \theta \limply \Inth{n+1})}}
    }{
      \lsequent{\theta \ge 0, \,  \dbox{\act_i}{\Reach{\theta}{\theta+T}}}{{\dbox{\alpha_i}{\Inth{n+1}}}}
    }
  }{
    \lsequent{\lor_i \, \exists \theta\!\ge\!0 \, \dbox{\act_i}{\Reach{\theta}{\theta+T}}}{\lor_i \, \dbox{\alpha_i}{\Inth{n+1}}}
  }
}{
  \lsequent{\dbox{\step}{\Inth{n}}}{\lor_i \, \dbox{\alpha_i}{\Inth{n+1}}}
}
\end{sequentdeduction}
where we abbreviate $\Gamma \define \theta \ge 0, \, \dbox{\act_i}{\Reach{\theta}{\theta+T}}$. In the case where $t \ge \theta$ after a control cycle, the agent has reached $\Inth{n}$ and therefore $\Inth{n+1}$:
\begin{sequentdeduction}[array]
\linfer[composeb]{
  \linfer[]{
    \linfer[]{
      \lclose
    }{
      \lsequent{\Gamma}{\dbox{\act_i}{R_i(\theta, \theta+T)}}
    }
  }{
    \lsequent{\Gamma}{\dbox{\act_i}{\dbox{\plant_{\theta+T}}{(\safe \land (t \ge \theta \limply I^{n}))}}}
  }
}{
  \lsequent{\Gamma}{\dbox{\act_i \seq \plant}{(t \ge \theta \limply \Inth{n+1})}}
}
\end{sequentdeduction}
In the case where $t \le \theta$ after a control cycle, the agent must perform the same action again with a timeout of $\theta - t$.
\begin{sequentdeduction}[array]
\linfer[]{
  \linfer[]{
    \linfer[]{
      \linfer[]{
        \linfer[]{ \lclose}
        {\linfer[]{
          \lsequent{\Gamma}{\dbox{\act_i}{\Reach{\theta}{\theta+T}}}
        }{
          \lsequent{\Gamma}{\dbox{\act_i \seq \plant}{\Reach{\theta-t}{\theta-t+T}}}
        }}
      }{
        \lsequent{\Gamma}{\dbox{\act_i \seq \plant \seq \act_i}{\Reach{\theta-t}{\theta-t+T}}}
      }
    }{
      \lsequent{\Gamma}{\dbox{\alpha_i}{(t \le \theta \limply \dbox{\act_i}{\Reach{\theta-t}{\theta-t+T}})}}
    }
  }{
    \lsequent{\Gamma}{\dbox{\alpha_i}{(t \le \theta \limply (\exists \rho\!\le\!0\, \dbox{\act_i}{\Reach{\rho}{\rho+T}}))}}
  }
}{
  \lsequent{\Gamma}{\dbox{\alpha_i}{(t \le \theta \limply \Inth{n+1})}}
}
\end{sequentdeduction}
This concludes the proof. \qed
\end{proofE}

\begin{figure}[htb!]
  \centering
  \includegraphics[width=0.85\textwidth]{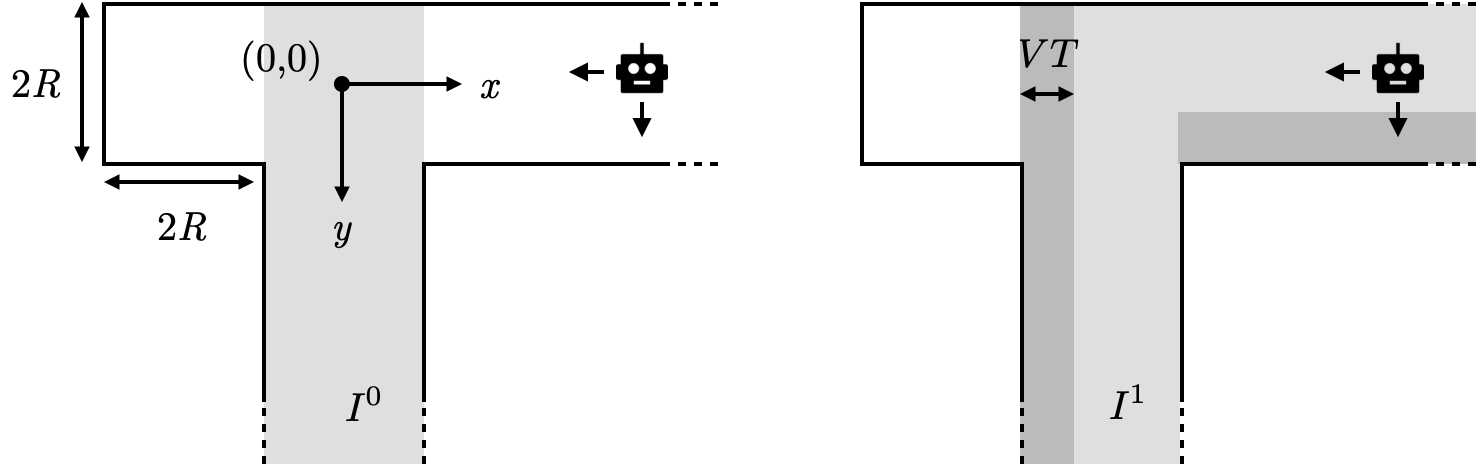}
  \caption{Robot navigating a corridor (\rref{model:corridor}). A 2D robot must navigate safely within a corridor with a dead-end without crashing against a wall. The corridor extends infinitely on the bottom and on the right. The robot can choose between going left and going down with a constant speed $V$. The left diagram shows $I^0$ in gray. The right diagram shows $I^1$ under the additional assumption $VT < 2R$ ($I^1$ and $I^0$ are otherwise equivalent). A darker shade of gray is used for regions of $I^1$ where only one of the two available actions is safe according to $G^1$.}\label{fig:corridor}
\end{figure}

\rref{thm:sn-solution} establishes a nontrivial result since it overcomes the significant gap between the \emph{fantasized} game that defines $I^n$ and the \emph{real} game being played by a time-triggered controller.
\tacastext{%
The proof critically relies on the action permanence assumption along with a result \cite[Lemma 6]{arxiv} establishing that ODEs preserve a specific form of reach-avoid property as a result of being deterministic.
}{%
Our proof critically relies on the action permanence assumption along with the following property of differential equations, which establishes that ODE programs preserve a specific form of reach-avoid property as a result of being deterministic.

\begin{lemmaE}[][restate]\label{lem:ode-reachability}
  Consider a property of the form $\Reach{a}{b} \!\define\! \dbox{\alpha_b}{(S \land (t \ge a \limply I))}$ with $\alpha_b \!\define\! (\pumod{t}{0} \seq \curly{\pevolvein{x'=f(x), t'=1}{Q \land t \le b}})$.
  Then this formula is valid:
  \[ {c \le b \land \Reach{a}{b} \limply \dbox{\alpha_c}{\Reach{a-t}{b-t}}}. \]
\end{lemmaE}
\begin{proofE}
  This follows from the semantics of $\dL$ since   all involved differential equations are the same and
$t\leq c\leq b$ is the duration that passes during $\alpha_c$, and thus explaining the offset of $-t$ on the time interval arguments of $\Reach{a}{b}$
  \qed.
\end{proofE}}
\paragraph{Example.} As an illustration, consider the example in \rref{fig:corridor} and \rref{model:corridor} of a 2D robot moving in a corridor that forms an angle.
The robot is only allowed to move left or down at a constant velocity and must not crash against a wall. Computing $\Inth{0}$ gives us the vertical section of the corridor, in which going down is a safe one-step fallback. Computing $\Inth{1}$ forces us to distinguish two cases. If the corridor is wider than the maximal distance travelled by the robot in a control cycle ($VT > 2R$), then the upper section of the corridor is controllable (with the exception of a dead-end that we prove to be uncontrollable in \rref{sec:optimality}). On the other hand, if the corridor is too narrow, then $\Inth{1}$ is equivalent to $\Inth{0}$. Formally, we have $I^1 \Define (y>-R \,\land\, \abs{x} < R) \ \lor \ (VT<2R \,\land\, (x>-R \,\land\, \abs{y} < R)).$
Moreover, computing $I^2$ gives a result that is equivalent to $I^1$. From this, we can conclude that $I^1$ is equivalent to $I^n$ for all $n \ge 1$. Intuitively, it is optimal with respect to \emph{any} finite fallback strategy (restricted to permanent actions).

\begin{model}[htb]
  \small
  \setcounter{modelline}{0}
  \caption{Robot navigating a corridor with framed solutions of holes.}\label{model:corridor}%
  \begin{spreadlines}{0.3ex}%
  \begin{align*}
    \text{\assum} &\,\big|\mline{line:corridor-assums} V > 0 \,\land\, T > 0  \\
    \text{\ctrlable} &\,\big|\mline{line:corridor-init} \land \, \highlightSynth{(y>-R \,\land\, \abs{x} < R) \lor (VT<2R \,\land\, (x>-R \,\land\, \abs{y} < R))} \limply [\{ \\
    \text{\ctrl} &\left|
    \begin{aligned}
      &\mline{line:corridor-left} \qquad ( \quad \parens{\ptest{\highlightSynth{x > -R + VT}} \seq \pumod{v_x}{-V} \seq \pumod{v_y}{0}} \\
      &\mline{line:corridor-down} \qquad \cup \  (\ptest{\highlightSynth{y<R-VT \,\lor\, x < R}} \seq \pumod{v_x}{0} \seq \pumod{v_y}{V}) \quad ) \seq \\
    \end{aligned}
    \right.\\
    \text{\plant} &\,\big|\mline{line:corridor-plant} \qquad (\pumod{t}{0} \seq \{x'=v_x, y'=v_y, t'=1 \ \& \ t \le T\}) \\
    \text{\safe} &\,\big|\mline{line:corridor-safety} \}^\ast] ((x > -3R \,\land\, \abs{y} < R) \lor (y > -R \,\land\, \abs{x} < R))
  \end{align*}
  \end{spreadlines}
\end{model}

The controllable invariant unrolling $\Inth{n}$ has a natural stopping criterion.

\begin{lemmaE}
  If \(\Inth{n} \lequiv \Inth{n+1}\) is valid for some $n \ge 0$,
  then \(\Inth{n} \lequiv \Inth{m}\) is valid for all $m \ge n$
  and \(\Inth{n} \lequiv \Inth{\omega}\) is valid
  where \(\Inth{\omega} \define \dbox{\drepeat{\step} \seq \forever}{\safe}\).
\end{lemmaE}
\begin{proofE}
  The first part is simply a case of a recursive sequence $\Inth{n+1} \define F(\Inth{n})$ reaching a fixpoint ($F(I) \define I \lor \dbox{\step}{I}$).
  Let us then prove the $\entails \Inth{n} \lequiv \Inth{\omega}$ equivalence, or rather the nontrivial direction $\entails \Inth{\omega} \limply \Inth{n}$. From $\entails \Inth{n} \lequiv \Inth{n+1}$, we get $\entails \Inth{n} \lequiv \Inth{n} \lor \dbox{\step}{\Inth{n}}$ and so $\entails \dbox{\step}{\Inth{n}} \limply \Inth{n}$. In addition, by the monotonicity of $(\Inth{n})_n$, we have $\entails \Inth{0} \limply \Inth{n}$. The rest follows from the \irref{dFP} rule:

  \begin{sequentdeduction}[array]
    \linfer[]
    {
      \linfer[dFP]
      {
        \linfer[orl]
        {
          \lsequent{\Inth{0}} {\Inth{n}}!
          \lsequent{\dbox{\step}{\Inth{n}}} {\Inth{n}}
        }
        {
          \lsequent{\Inth{0} \lor \dbox{\step}{\Inth{n}}} {\Inth{n}}
        }
      }{
        \lsequent{\dbox{\drepeat{\step}}{\Inth{0}}} {\Inth{n}}
      }
    }
    {
      \lsequent{\Inth{\omega}} {\Inth{n}}
    }
  \end{sequentdeduction}
  \qed
\end{proofE}


\subsection{Proving Optimality via the Dual Game}\label{sec:optimality}

Suppose one found a controllable invariant $I$ using techniques from the previous section. To prove it optimal, one must show that $\entails\,\assum \limply (\Iopt \limply I)$.
By contraposition and \(\dbox{\alpha}{P} \lbisubjunct \neg\ddiamond{\alpha}{\neg P}\) (\irref{box}), this is equivalent to proving that:
\begin{equation}\label{eq:optimality}
  \entails \ \assum \land \neg I \limply \underbrace{\ddiamond{\prepeat{((\cap_i \, \act_i) \seq \plant)}}{\neg \safe}}_{\neg \Iopt}.
\end{equation}
 We define the largest \emph{un}controllable region $\Uopt \equiv \neg \Iopt$ as the right-hand side of implication \ref{eq:optimality} above.
 Intuitively, $\Uopt$ characterizes the set of all states from which the environment (played by Angel) has a winning strategy against the controller (played by Demon) for reaching an unsafe state.
 In order to prove the optimality of $I$, we compute a sequence of increasingly strong approximations $U$ of $\Uopt$ such that $U \limply \Uopt$ is valid. We do so via an iterative process, in the spirit of how we approximate $\Iopt$ via bounded fallback unrolling (\rref{sec:unrolling}), although the process can be guided by the knowledge of $I$ this time. If at any point we manage to prove that $\assum \limply (I \lor U)$ is valid, then $I$ is optimal.

One natural way to compute increasingly good approximations of $\Uopt$ is via loop unrolling.
The idea is to improve approximation $U$ by adding states from where the environment can reach $U$ by running the control loop once, formally,
$\ddiamond{(\cap_i \, \act_i) \seq \plant}{U}$.
This unrolling principle can be useful.
However, it only augments $U$ with new states that can reach $U$ in time $T$ at most.
So it cannot alone prove optimality in cases where violating safety from an unsafe state takes an unbounded amount of time.

For concreteness, let us prove the optimality of $\Inth{0}$ in the case of \rref{model:etcs}. In~\cite{DBLP:conf/icfem/PlatzerQ09} essentially the following statement is proved when arguing for optimality:
\( \ \entails \assum \land \neg I^0 \limply \ddiamond{(\pumod{a}{-B} \seq \plant)^*} \neg \safe. \)
This is identical to our optimality criterion from \rref{eq:optimality}, except that Demon's actions are restricted to braking. Intuitively, this restriction is sound since accelerating always makes things worse as far as safety is concerned. If the train cannot be saved with braking alone, adding the option to accelerate will not help a bit.
In this work, we propose a method for formalizing such arguments within $\dGL$ to arbitrary systems.

Our idea for doing so is to consider a system made of two separate copies of our model. One copy has all actions available whereas the other is only allowed a single action (e.g. braking). Given a safety metric $m$ (i.e. a term $m$ such that $\entails m \le 0 \limply \neg \safe$), we can then formalize the idea that ``action $i$ is always better w.r.t safety metric $m$" within this joint system.

\begin{definition}[Uniform Action Optimality]
  Consider a finite number of discrete \dL programs $\alpha_i$ and $p \equiv \{x'=f(x) \ \& \ Q\}$.
  Let $V = \BV(p) \cup \cupfold_i \BV(\alpha_i)$ be the set of all variables written by $p$ or some $\alpha_i$.
  For any term $\theta$ and integer $n$, write $\tagged{\theta}{n}$ for the term that results from $\theta$ by renaming all variables $v \in V$ to a fresh tagged version $\tagged{x}{n}$.
  Using a similar notation for programs and formulas, define $\tagged{p}{1,2} \equiv \{(\tagged{x}{1})' = f(\tagged{x}{1}), (\tagged{x}{2})' = f(\tagged{x}{2}) \ \& \ \tagged{Q}{1} \wedge \tagged{Q}{2} \}$.
  We say that action $j$ is \emph{uniformly optimal} with respect to safety metric $m$ if and only if:
  \[\entails \ \tagged{m}{1} \ge \tagged{m}{2} \limply \dbox{\tagged{\alpha_j}{1} \seq (\cup_i \, \tagged{\alpha_i}{2}) \seq \tagged{p}{1,2}}{\tagged{m}{1} \ge \tagged{m}{2}}. \]
  $\UniformlyOptimal{j}{(\alpha_i)_i}{p}{m}$ denotes that action $j$ is uniformly optimal with respect to $m$ for actions $\alpha_i$ and dynamics $p$.
\end{definition}
With such a concept in hand, we can formally establish the fact that criterion \rref{eq:optimality} can be relaxed in the existence of uniformly optimal actions.
\begin{theoremE}\label{thm:ghost-player}
Consider a finite number of discrete \dL programs $\alpha_i$ such that $\entails \ddiamond{\alpha_i} \true$ for all $i$ and $p \equiv \{x'=f(x) \ \& \ q \ge 0\}$.
Then, provided that $\UniformlyOptimal{j}{(\alpha_i)_i}{p}{m}$ and $\UniformlyOptimal{j}{(\alpha_i)_i}{p}{-q}$ (no other action stops earlier because of the domain constraint), we have:
\[\entails \ \ddiamond{((\cap\, \alpha_i) \seq p)^*}{m \le 0} \lbisubjunct \ddiamond{(\alpha_j \seq p)^*}{m \le 0} \enspace .  \]
\end{theoremE}
\begin{proofE}
The nontrivial implication to prove is:
\[\entails \ddiamond{(\alpha_j \seq p)^*}{m \le 0} \ \limply \ \ddiamond{((\cap\, \alpha_i) \seq p)^*}{m \le 0}. \]
We do so by proving:
\begin{equation}\label{eq:ghost-chain} \Gamma \entails \ddiamond{(\tagged{\alpha_j}{1} \seq \tagged{p}{1})^*}{\tagged{m}{1} \le 0} \ \limply \  \ddiamond{((\cap_i \tagged{\alpha_i}{2}) \seq \tagged{p}{2})^*}{\tagged{m}{2} \le 0,} \end{equation} where $\Gamma \equiv \landfold_{x \in V} (\tagged{x}{1} = \tagged{x}{2})$ and $V \equiv \BV(p) \cup \cupfold_i \BV(\alpha_i)$.
To prove \rref{eq:ghost-chain}, we chain together three implications:

\begin{enumerate}
  \item $\Gamma \entails \ddiamond{(\tagged{\alpha_j}{1} \seq \tagged{p}{1})^*}{\tagged{m}{1} \le 0} \ \limply \ \ddiamond{(\tagged{\alpha_j}{1} \seq (\cap_i \tagged{\alpha_i}{2}) \seq \tagged{p}{1,2})^*}{\tagged{m}{1} \le 0}$:
  \begin{enumerate}
    \item To prove the implication above, we consider a sequence of states $s_1 \dots s_n$ such that $s_1 \in \Sem{\Gamma}$, $s_n \in \Sem{\tagged{m}{1} \le 0}$ and $(s_i, s_{i+1} \in \Sem{\tagged{\alpha_j}{1} \seq \tagged{p}{1}})$ for all $i$. We must prove that $s_1 \in \ddiamond{(\tagged{\alpha_j}{1} \seq (\cap_i \tagged{\alpha_i}{2}) \seq \tagged{p}{1,2})^*}{\tagged{m}{1} \le 0}$.
    \item We say that two states $s$ and $s'$ are \emph{1-equivalent} (written $s \sim_{(1)} s'$) if they only differ on variables tagged with 2. Using this definition, it is enough to prove the following fact: for all $i \le n$ and any state $s$ such that $s \sim_{(1)} s_i$ and $s \in \Sem{\tagged{q}{1} \le \tagged{q}{2}}$, we have $s \in \ddiamond{(\tagged{\alpha_j}{1} \seq (\cap_i \tagged{\alpha_i}{2}) \seq \tagged{p}{1,2})^*}{\tagged{m}{1} \le 0}$.
    \item We prove the fact above by descending induction on $i$. The base case for $i=n$ follows from our assumption on $s_n$. The inductive case considers a state $s$ such that $s \sim_{(1)} s_{i-1}$ and $s \in \Sem{\tagged{q}{1} \le \tagged{q}{2}}$. By assumption, $(s_{i-1}, s_i) \in \Sem{\tagged{\alpha_j}{1} \seq \tagged{p}{1}}$. Thus, there exists $s'$ such that $(s, s') \in \Sem{\tagged{\alpha_j}{1} \seq (\cap_i \tagged{\alpha_i}{2}) \seq \tagged{p}{1,2}}$,  $s' \sim_{(1)} s_i$ and $s' \in \Sem{\tagged{q}{1} \le \tagged{q}{2}}$. Note that the assumption that $\UniformlyOptimal{j}{(\alpha_i)_i}{p}{-q}$ is critical in establishing the existence of $s'$, by ensuring that $\tagged{p}{1,2}$ can be run for at least as long $\tagged{p}{1}$. We conclude by using the induction hypothesis on $s'$.
  \end{enumerate}
  \item $\Gamma \entails \ddiamond{\gamma}{\tagged{m}{1} \le 0} \limply \ddiamond{\gamma}{\tagged{m}{2} \le 0}$ with $\gamma \equiv (\tagged{\alpha_j}{1} \seq (\cap_i \tagged{\alpha_i}{2}) \seq \tagged{p}{1,2})^*$:
  \begin{enumerate}
    \item The key fact we are using here is that for any game $\beta$ and formulas $P, Q$, we have $\entails \dbox{\dfree{\beta}}{(P \limply Q)} \limply \ddiamond{\beta}{P} \limply \ddiamond{\beta}{Q}$ where $\dfree{\beta}$ is obtained from $\beta$ by removing all applications of the dual operator. Intuitively, this is true since $\dbox{\dfree{\beta}}{(P \limply Q)}$ ensures that $P \limply Q$ is true in every reachable game state, independently of both players' strategies.
    \item Per the theorem assumption, we have $\Gamma \entails \dbox{\dfree{\gamma}}{(\tagged{m}{1} \ge \tagged{m}{2})}$. Using the monotonicity rule \irref{M}, we obtain $\Gamma \entails \dbox{\dfree{\gamma}}{(P \limply Q)}$ with $P \equiv \tagged{m}{2} \le 0$ and $Q \equiv \tagged{m}{1} \le 0$. We can then conclude using the previous point.
  \end{enumerate}
  \item $\entails \ddiamond{(\tagged{\alpha_j}{1} \seq (\cap_i \tagged{\alpha_i}{2}) \seq \tagged{p}{1,2})^*}{\tagged{m}{2} \le 0} \ \limply\ \ddiamond{((\cap_i \tagged{\alpha_i}{2}) \seq \tagged{p}{2})^*}{\tagged{m}{2} \le 0}$:
   \begin{enumerate}
    \item This last implication can be proved in a similar way than  (1.), although the $\UniformlyOptimal{j}{(\alpha_i)_i}{p}{-q}$ assumption is not needed since we are removing ODE domain constraints instead of adding them.
  \end{enumerate}
\end{enumerate}
By chaining everything, we get:
\[ \Gamma \entails \ddiamond{(\tagged{\alpha_j}{1} \seq \tagged{p}{1})^*}{\tagged{m}{1} \le 0} \ \limply \  \ddiamond{((\cap_i \tagged{\alpha_i}{2}) \seq \tagged{p}{2})^*}{\tagged{m}{2} \le 0}, \] which closes the proof. \qed
\end{proofE}

\afterProofE A general heuristic for leveraging \rref{thm:ghost-player} to grow $U$ automatically works as follows. First, it considers $R \equiv \assum \land \neg I \land \neg U$ that characterizes states that are not known to be controllable or uncontrollable. Then, it picks a disjunct $\landfold_j R_{j}$ of the disjunctive normal form of $R$ and computes a forward invariant region $V$ that intersects with it: $V \equiv \landfold_j \{ R_{j} : \assum, \, R_{j} \vdash \dbox{(\cup_i \, \act_i) \seq \plant}{R_{j}} \}$. Using $V$ as an assumption to simplify $\neg U$ may suggest metrics to be used with \rref{thm:ghost-player}. For example, observing $\entails V \limply(\neg U \limply (\theta_1 > 0 \land \theta_2 > 0))$ suggests picking metric $m \equiv \min(\theta_1, \theta_2)$ and testing whether $\UniformlyOptimal{j}{\act}{p}{m}$ is true for some action $j$. If such a uniformly optimal action exists, then $U$ can be updated as $U \gets U \,\lor\, (V \land \ddiamond{(\act_j \seq \plant)^*}{m \le 0})$. The solution $\Inth{1}$ for the corridor (\rref{model:corridor}) can be proved optimal automatically using this heuristic in combination with loop unrolling.


\subsection{Implementing the Reduction Oracle}\label{sec:reduce}

The CESAR algorithm assumes the existence of a \emph{reduction oracle} that takes as an input a loop-free $\dGL$ formula and attempts to compute an equivalent formula within the fragment of propositional arithmetic. When an exact solution cannot be found, an implicant is returned instead and flagged appropriately (\rref{def:reduce}). This section discusses our implementation of such an oracle.

As discussed in \rref{sec:single-action}, exact solutions can be  computed systematically when all ODEs are solvable by first using the $\dGL$ axioms to eliminate modalities \tacastext{}{(see \rref{app:proofs})} and then passing the result to a \emph{quantifier elimination algorithm} for first-order arithmetic~\cite{caviness2012quantifier,tarski}.
Although straightforward in theory, a na\"ive implementation of this idea hits two practical barriers.
First, quantifier elimination is expensive and its cost increases rapidly with formula complexity~\cite{DBLP:journals/jsc/DavenportH88,DBLP:journals/jsc/Weispfenning88}.
Second, the output of existing QE implementations can be unnecessarily large and redundant. In iterated calls to the reduction oracle, these problems can compound each other.

To alleviate this issue, our implementation performs \emph{eager simplification} at intermediate stages of computation, between some axiom application and quantifier-elimination steps.
This optimization significantly reduces output solution size and allows CESAR to solve a benchmark that would otherwise timeout after 20 minutes in 26s.
\tacastext{\cite[Appendix E]{arxiv}}{\rref{app:reduce-example}} further discusses the impact of eager simplification.
Still, the doubly exponential complexity of quantifier elimination puts a limit on the complexity of problems that \AlgoName can currently tackle.

In the general case, when ODEs are not solvable, our reduction oracle is still often able to produce \textit{approximate} solutions using differential invariants generated automatically by existing tools \cite{DBLP:journals/fmsd/SogokonMTCP22}.
Differential invariants are formulas that stay true throughout the evolution of an ODE system.
\footnote{\dGL provides ways to reason about differential invariants without solving the corresponding differential equation. For example, for an invariant of the form $e=0$, the differential invariant axiom is $\dbox{\{x'=f(x)\}}{e=0}\lbisubjunct (e=0 \land \dbox{\{x'=f(x)\}}{e'=0})$.}
To see how they apply, consider the case of computing $\reduce(\dbox{\{x'=f(x)\}}{P}, A)$ where $P$ is the postcondition formula that must be true after executing the differential equation, and $A$ is the assumptions holding true initially.
Suppose that formula $D(x)$ is a differential invariant such that $D(x)\limply P$ is valid.
Then, a precondition sufficient to ensure that $P$ holds after evolution is $A \limply D(x)$.
\tacastext{%
For example, to compute the precondition for the dynamics of the \texttt{parachute} benchmark, our reduction oracle first uses the Pegasus tool \cite{DBLP:journals/fmsd/SogokonMTCP22} to identify a Darboux polynomial, suggesting an initial differential invariant $D_0$.%
}{%
For a concrete example, \rref{app:parachute-ode} shows how our reduction oracle computes the precondition for the dynamics of the \texttt{parachute} benchmark.
It first uses the Pegasus tool \cite{DBLP:journals/fmsd/SogokonMTCP22} to identify a Darboux polynomial, suggesting an initial differential invariant $D_0$.}
Once we have $D_0$, the additional information required to conclude post condition $P$ is $D_0 \limply P$.
To get an invariant formula that implies $D_0 \limply P$, eliminate all the changing variables $\{x, v\}$ in the formula $\forall x \, \forall v\ (D_0 \limply P)$, resulting in a formula $D_1$.
$D_1$ is a differential invariant since it features no variable that is updated by the ODEs.
Our reduction oracle returns $D_0 \land D_1$, an invariant that entails postcondition $P$.
\tacastext{}{More details on our implementation of \reduce and how it deals with ODEs in particular can be found in \rref{app:reduce}.}


\subsection{The \AlgoName Algorithm}\label{sec:algo}

\newcommand{\ValidCheckName}{\textsf{valid}}
\newcommand{\ValidCheck}[2]{\ValidCheckName(#1, \ #2)}
\newcommand{\exactTmp}{e}
\newcommand{\exactI}{e_{I}}
\newcommand{\exactG}{e_{G}}
\newcommand{\Yield}{\textbf{yield}\ }
\newcommand{\DiamondTest}[1]{\textsf{optimal}({#1})}

\newcommand{\AlgoAnd}[2]{#1 \textbf{ and } #2}
\newcommand{\AlgoOr}[2]{#1 \textbf{ or } #2}

\newcommand{\guarantee}[1]{``\emph{#1}"}

\algnewcommand\algorithmicforeach{\textbf{for each}}
\algdef{S}[FOR]{ForEach}[1]{\algorithmicforeach\ #1\ \algorithmicdo}

The \AlgoName algorithm for synthesizing control envelopes is summarized in \rref{alg:main}. It is expressed as a generator that yields a sequence of solutions with associated optimality guarantees.
Possible guarantees include \guarantee{sound} (no optimality guarantee, only soundness), \guarantee{$k$-optimal} (sound and optimal w.r.t all $k$-switching fallbacks with permanent actions), \guarantee{$\omega$-optimal} (sound and optimal w.r.t all finite fallbacks with permanent actions) and \guarantee{optimal} (sound and equivalent to $\Sopt$).
\rref{line:diamond-test} performs the optimality test described in \rref{sec:optimality}. Finally, \rref{line:soundness-check} performs an important soundness check for the cases where an approximation has been made along the way of computing $(\Inth{n}, \Gnth{n})$.
In such cases, $I$ is not guaranteed to be a controllable invariant and thus \rref{case:some-act} of \rref{def:synthesis-problem} must be checked explicitly.

When given a problem with solvable ODEs and provided with a complete QE implementation within \reduce, \AlgoName is guaranteed to generate a solution in finite time with an \guarantee{$n$-optimal} guarantee at least ($n$ being the unrolling limit).

\begin{algorithm}
  \caption{\AlgoName: Control Envelope Synthesis via Angelic Refinements}\label{alg:main}
  \begin{algorithmic}[1]
    \State \textbf{Input:} a synthesis problem (as defined in \rref{sec:problem}), an unrolling limit $n$.
    \State \textbf{Remark:} $\ValidCheckName$ is defined as $\ValidCheck{F}{A} \equiv (\fst(\reduce(\neg F, A)) = \false)$.
    \State $k \gets 0$
    \State $I, \exactI \gets \reduce(\dbox{\forever}{\safe}, \ \assum)$
      \While{$k \le n$}
      \State $\exactG \gets \true$ 
      \ForEach{i}
        \State $G_i, \exactTmp \gets \reduce(\dbox{\act_i \seq \plant}{I}, \ \assum)$
        \State $\exactG \gets \AlgoAnd{\exactG}{\exactTmp}$
      \EndFor
      \If{$\AlgoOr{(\AlgoAnd{\exactG}{\exactI})}{\ValidCheck{I \limply \lorfold_i G_i}{\assum}}$} \label{line:soundness-check}
        \If{$\AlgoAnd{\exactG}{\DiamondTest{I}}$} \label{line:diamond-test}
          \State \Yield $((I, G), \text{``optimal"})$
          \State \Return
        \ElsIf{$\AlgoAnd{\exactG}{\exactI}$}  \Yield $((I, G), \text{``$k$-optimal"})$
        \Else \ \Yield $((I, G), \text{``sound"})$ 
        \EndIf
      \EndIf
      \State $I', \exactTmp \gets \reduce(I \lor \dbox{\step}{I}, \ \assum)$
      \State $\exactI \gets \AlgoAnd{\exactI}{\exactTmp}$
      \If{$\AlgoAnd{\AlgoAnd{\exactG}{\exactI}}{\ValidCheck{I' \limply I}{\assum}}$}
        \State \Yield $((I, G), \text{``$\omega$-optimal"})$ 
        \State \Return
      \EndIf
      \State $I \gets I'$
      \State $k \gets k + 1$
    \EndWhile
  \end{algorithmic}
\end{algorithm}


\section{Benchmarks and Evaluation}
\label{sec:evaluation}

To evaluate our approach to the \synthesis problem, we curate a benchmark suite with diverse optimal control strategies.
As \rref{tab:results} summarizes, some benchmarks have non-solvable dynamics, while others require a sequence of clever control actions to reach an optimal solution.
Some have \textit{state-dependent fallbacks} where the current state of the system determines which action is ``safer'', and some are drawn from the literature.
We highlight a couple of benchmarks here.
See \tacastext{\cite[Appendix D]{arxiv}}{\rref{app:benchmarks}} for a discussion of the full suite and the synthesized results, and \cite{artifact} for the benchmark files and evaluation scripts.

\tacastext{\texttt{Power Station}}{\hyperref[app:power-station]{\texttt{Power Station}}} is an example where the optimal control strategy involves two switches, corresponding to two steps of unrolling.
A power station can either produce power or dispense it to meet a quota, but never give out more than it has produced.
Charging is the fallback action that is safe for all time \textit{after} the station has dispensed enough power.
However, to cover all controllable states, we need to switch at least two times, so that the power station has a chance to produce energy and then dispense it, before settling back on the safe fallback.
\tacastext{\texttt{Parachute}}{\hyperref[app:parachute]{\texttt{Parachute}}} is an example of a benchmark with non-solvable, hyperbolic dynamics.
A person jumps off a plane and can make an irreversible choice to open their parachute.
The objective is to stay within a maximum speed that is greater than the terminal velocity when the parachute is open.

We implement \AlgoName in Scala, using Mathematica for simplification and quantifier elimination, and evaluate it on the benchmarks.
Simplification is an art \cite{simplification,knuth}.
We implement additional simplifiers with the \texttt{Egg} library \cite{egg} and SMT solver \texttt{z3} \cite{DBLP:journals/corr/DeMouraM08}.
Experiments were run on a 32GB RAM M2 MacBook Pro machine.
\AlgoName execution times average over 5 runs.

\AlgoName synthesis is automatic.
The optimality tests were computed manually.
\rref{tab:results} summarizes the result of running \AlgoName.
Despite a variety of different control challenges, \AlgoName is able to synthesize safe and in some cases also optimal safe control envelopes within a few minutes.
As an extra step of validation, synthesized solutions are checked by the hybrid system theorem prover \KeYmaeraX \cite{DBLP:conf/cade/FultonMQVP15}.
All solutions are proved correct, with verification time as reported in the last column of \rref{tab:results}.

\begin{table}[htbp]
\caption{Summary of \AlgoName experimental results}
\label{tab:results}
\centering
\begin{tabular}{lrrcccc}
  \toprule
  \thead{Benchmark} & \thead{Synthesis\\Time (s)} & \thead{Checking\\Time (s)} & \thead{Optimal}  & \thead{Needs\\Unrolling} & \thead{Non\\Solvable\\Dynamics} \\
  \midrule
  \tacastext{ETCS Train}{\hyperref[app:etcs]{ETCS Train}} \cite{DBLP:conf/icfem/PlatzerQ09}   & 14  & 9   & \checkmark        &            &             \\
  \tacastext{Sled}{\hyperref[app:sled]{Sled}}                                                 & 20  & 8   & \checkmark        &            &             \\
  \tacastext{Intersection}{\hyperref[app:intersection]{Intersection}}                         & 49  & 44  & \checkmark        &            &             \\
  \tacastext{Parachute}{\hyperref[app:parachute]{Parachute}} \cite{DBLP:conf/itp/FultonMBP17} & 46  & 8   &                   &            &  \checkmark \\
  \tacastext{Curvebot}{\hyperref[app:curvebot]{Curvebot}}                                     & 26  & 9   &                   &            &  \checkmark \\
  \tacastext{Coolant}{\hyperref[app:coolant]{Coolant}}                                        & 49  & 20  & \checkmark        & \checkmark &             \\
  \tacastext{Corridor}{\hyperref[app:corridor]{Corridor}}                                     & 20  & 8   & \checkmark        & \checkmark &             \\
  \tacastext{Power Station}{\hyperref[app:power-station]{Power Station}}                      & 26  & 17  & \checkmark        & \checkmark &             \\
  \bottomrule
\end{tabular}
\end{table}
%


\section{Related Work}
\label{sec:related}
\paragraph{Hybrid controller synthesis} has received significant attention \cite{LIU202230,Tabuada09,Belta17}, with popular approaches using temporal logic \cite{525480,Belta17,chen20}, games \cite{DBLP:conf/cdc/NerodeY92,DBLP:journals/IEEE/TomlinLS00}, and CEGIS-like guidance from counterexamples \cite{DBLP:journals/sttt/Solar-Lezama13,DBLP:journals/acta/AbateBCDKKP20,DBLP:conf/emsoft/RavanbakhshS16,Dai2020CounterexampleGS}.
\AlgoName, however, solves the different problem of synthesizing control \emph{envelopes} that strive to represent not one but \emph{all} safe controllers of a system.
Generating \emph{valid} solutions is not an issue (a trivial solution always exists that has an empty controllable set).
The real challenge is \emph{optimality} which imposes a higher order constraint because it reasons about the relationship between possible valid solutions, and cannot, e.g., fit in the CEGIS quantifier alternation pattern $\exists\forall$.
So simply adapting existing controller synthesis techniques does not solve symbolic control envelope synthesis.

\emph{Safety shields} computed by numerical methods \cite{10.1609/aaai.v32i1.11797,10.1109/tac.2018.2876389,kochenderfer2012next} serve a similar function to our \emph{control envelopes} and can handle dynamical systems that are hard to analyze symbolically.
However, they scale poorly with dimensionality and do not provide rigorous formal guarantees due to the need of discretizing continuous systems.
Compared to our symbolic approach, they cannot handle unbounded state spaces (e.g. our infinite corridor) nor produce shields that are parametric in the model's parameters without hopelessly increasing dimensionality.

On the optimality side, a systematic but manual process was used to design a safe European Train Control System (ETCS) and justify it as optimal with respect to specific train criteria \cite{DBLP:conf/icfem/PlatzerQ09}.
Our work provides the formal argument filling the gap between such case-specific criteria and end-to-end optimality.
\AlgoName is more general and automatic.


\section{Conclusion}
\label{sec:conclusion}

This paper presents the \AlgoName algorithm for \AlgoLong.
It is the first approach to automatically synthesize symbolic control envelopes for hybrid systems.
The synthesis problem and optimal solution are characterized in differential game logic.
Through successive refinements, the optimal solution in game logic is translated into a controllable invariant and control conditions.
The translation preserves safety.
For the many cases where refinement additionally preserves optimality, an algorithm to test optimality of the result post translation is presented.
The synthesis experiments on a benchmark suite of diverse control problems demonstrate {\AlgoName}'s versatility.
For future work, we plan to extend to additional control shapes, and to exploit the synthesized safe control envelopes for reinforcement learning.


\renewcommand{\doi}[1]{doi: \href{https://doi.org/#1}{\nolinkurl{#1}}}
\bibliographystyle{splncs04}
\bibliography{refs}

\tacastext{}{
\appendix
\clearpage


\section{Reduce Operation}\label{app:reduce}
We define \reduce after first introducing two helper functions that it requires.
Function \(\abductp(a, b)\) attempts to simplify \fol formula $a$ to \props assuming that $b$ holds. 
The second helper function, \odereduce (\rref{def:odereduce}), isolates the action of reduce on differential equations.
Since it is solely continuous programs that could lead to reduce failing to produce an exact solution, the \exact bit of \reduce depends on \odereduce.
\rref{fig:reduce} shows the definition of \reduce, eliding the \exact bit, which is simply true if all of the \odereduce calls that \reduce makes are exact, and false otherwise.

\begin{definition}[ODE reduction]\label{def:odereduce}
  Let \(\dmodality{\alpha}{} \in \{\langle \alpha \rangle, [\alpha]\}\), \(\bowtie \in \{\lbisubjunct,\limply\}\), and \(A\), \(Q\), and \(P\) be formulas in quantifier-free real arithmetic.
  An \emph{ODE reduction oracle \odereduce} is a function
  such that \[A \limply (\text{\emph{\odereduce}}(\dmodality{\{x' =f(x) \& Q\}}{}P, A) \bowtie \dmodality{\{x'=f(x) \& Q\}}{P})\] is valid.
  When \(\bowtie\) is \(\lbisubjunct\) then \emph{\odereduce} is exact, otherwise it is approximating.
\end{definition}
\begin{figure}[tbhp]
\begin{align*}
  \text{Let }\dmodality{\alpha}{} & \in \{\langle \alpha \rangle, [\alpha]\}\\
  \reduce(\dmodality{\alpha ; \beta}{} P, A) &= \reduce(\dmodality{\alpha}{}(\reduce(\dmodality{\beta}{P}, \top)), A) \\
  \reduce([\alpha \cup \beta]P, A) &= \reduce([\alpha]P, A) \land \reduce([\beta]P, A)\\
  \reduce(\langle \alpha \cup \beta \rangle P, A) &= \reduce(\langle \alpha \rangle P, A) \lor \reduce(\langle \beta \rangle P, A)\\
  \reduce(\dmodality{x:=e}{P}, A) &= \abductp(P\{e/x\}, A)\\
  \reduce([\ptest{f}]P, A) = \abductp(f \limply P, A) &
  \quad \reduce(\langle \ptest{f} \rangle P, A) = \abductp(f \land P, A)\\
  \reduce([\{x'=f(x) \& Q\}]P, A) &= \abductp (\odereduce([{x'=f(x) \& Q}]P,A), \top)\\
  \reduce(\langle\{x'=f(x) \& Q\}\rangle P, A) &= \abductp (\odereduce(\langle \{x'=f(x) \& Q\}\rangle P, A), \top)\\
  \reduce([\alpha^d]P) = \reduce(\langle \alpha \rangle P) &
  \quad \reduce(\langle \alpha^d \rangle P) = \reduce([\alpha]P)\\
  \reduce(P \land Q, A) = \abductp (P\land Q, A) &
  \quad \reduce(P \lor Q, A) = \abductp (P\lor Q, A) \\
  \reduce(P \limply Q, A) = \abductp (P\limply Q, A) &
  \quad \reduce(\neg P, A) = \abductp (\neg P, A)
\end{align*}
\caption{Definition of $\reduce$ (\exact elided). Notation $P\{e/x\}$ indicates $P$ with unbound occurrences of \(x\) replaced by expression $e$.
\odereduce isolates the effect of \reduce on ODEs.
\abductp simplifies and quantifier eliminates $P$ assuming \(A\).}
\label{fig:reduce}
\end{figure}
For solvable ODEs, \odereduce is implemented as an exact oracle in \rref{eq:ode-reduce-solvable}.
\begin{equation}
  \begin{aligned}
    \odereduce([\{x_1'=\theta_1,\cdots,x_n'=\theta_n \& Q\}]P, A) &=\\
    \forall t \bigl(\text{subst}(A,t) \land t\geq 0
     \land \forall 0{\le}s{\le}t\, & \text{subst}(Q, s)\bigr)
      \limply \text{subst}(P,t) \\
    \odereduce(\langle\{x_1'=\theta_1,\cdots,x_n'=\theta_n \& Q\}\rangle P, A) &=\\
    \exists t \bigl(\text{subst}(A,t) \land t\geq 0
      \land \forall 0{\le}s{\le}t\, & \text{subst}(Q, s)\bigr)
      \land \text{subst}(P,t) \\
      \text{ where } \text{subst}(f, t) = f\{\int_{x_i}^{t}\theta_i \cdot dt \, / \, x_i \},&\text{ $t$ fresh}, \dmodality{\alpha}{} \in \{\langle \alpha \rangle, [\alpha]\}
  \end{aligned}
  \label{eq:ode-reduce-solvable}
\end{equation}
In the general case, Pegasus \cite{DBLP:journals/fmsd/SogokonMTCP22}, a tool that automatically generates \textit{ODE invariants}, can often produce a formula satisfying the specification of \odereduce.
This may come at the cost of lost precision, possibly requiring \reduce to set \exact to false.

\begin{theoremE}[Correctness of \reduce]
  For any loop-free \dGL formula \(F\) and assumptions \(A \in \props\) the function \emph{\odereduce} either sets \exact{}=\(\ltrue\) and the formula
  \(A \limply (\reduce(F,A) \lbisubjunct F)\) is valid, or else it sets \exact{}=\(\lfalse\) and the formula
  \(A \limply (\reduce(F,A) \limply F)\) is valid.
\end{theoremE}
\begin{proofE}
  Follows from \dGL axioms being defined in terms of the decidable fragment of \fol, quantifier elimination being decidable, and the properties of \odereduce (\rref{def:odereduce}).
\end{proofE}

\section{Proofs}\label{app:proofs}

\subsection{Background}

The \dGL axioms and proof rules \cite{Platzer18} used here are summarized in \rref{fig:dGL-calculus}, noting that \(\phi\limply\psi\) and \(\lsequent{\phi}{\psi}\) have the same meaning.
The semantics \cite{Platzer18} is as follows.

{\renewcommand{\D}[2][]{\ifthenelse{\equal{#1}{}}{#2'}{\frac{d#2}{d{#1}}}}%

\begin{definition}[\dGL semantics] \label{def:dGL-semantics}
The \emph{semantics of a \dGL formula} $\phi$ is the subset \m{\imodel{\I}{\phi}\subseteq\linterpretations{\Sigma}{V}} of states in which $\phi$ is true.
It is defined inductively as follows
\begin{enumerate}
\item \(\imodel{\I}{p(\theta_1,\dots,\theta_k)} = \{\iportray{\I} \in \linterpretations{\Sigma}{V} \with (\ivaluation{\I}{\theta_1},\dots,\ivaluation{\I}{\theta_k})\in\iget[const]{\I}(p)\}\)
\item \(\imodel{\I}{\theta_1\sim\theta_2} = \{\iportray{\I} \in \linterpretations{\Sigma}{V} \with \ivaluation{\I}{\theta_1}\sim\ivaluation{\I}{\theta_2}\} \text{ where }{\sim}\in\{<,\leq,=,\geq,>\}\)
\item \(\imodel{\I}{\lnot\phi} = \scomplement{(\imodel{\I}{\phi})}\)
\item \(\imodel{\I}{\phi\land\psi} = \imodel{\I}{\phi} \cap \imodel{\I}{\psi}\)
\item
{\def\Im{\imodif[state]{\I}{x}{r}}%
\(\imodel{\I}{\lexists{x}{\phi}} =  \{\iportray{\I} \in \linterpretations{\Sigma}{V} \with \iget[state]{\Im} \in \imodel{\I}{\phi} ~\text{for some}~r\in\reals\}\)
}
\item \(\imodel{\I}{\ddiamond{\alpha}{\phi}} = \strategyfor[\alpha]{\imodel{\I}{\phi}}\)
\item \(\imodel{\I}{\dbox{\alpha}{\phi}} = \dstrategyfor[\alpha]{\imodel{\I}{\phi}}\)
\end{enumerate}
A \dGL formula $\phi$ is \emph{valid}, written \m{\entails{\phi}}, iff it is true in all states, i.e.\ \m{\imodel{\I}{\phi}=\linterpretations{\Sigma}{V}}.
\end{definition}

\begin{definition}[Semantics of hybrid games] \label{def:HG-semantics}
The \emph{semantics of a hybrid game} $\alpha$ is a function \m{\strategyfor[\alpha]{\cdot}} that, for each set of Angel's winning states \m{X\subseteq\linterpretations{\Sigma}{V}}, gives the \emph{winning region}, i.e.\ the set of states \m{\strategyfor[\alpha]{X}} from which Angel has a winning strategy to achieve $X$ in $\alpha$ (whatever strategy Demon chooses). It is defined inductively as follows
\begin{enumerate}
\item \(\strategyfor[\pupdate{\pumod{x}{\theta}}]{X} = \{\iportray{\I} \in \linterpretations{\Sigma}{V} \with \modif{\iget[state]{\I}}{x}{\ivaluation{\I}{\theta}} \in X\}\)
\item \(\strategyfor[\pevolvein{\D{x}=\genDE{x}}{\ivr}]{X} = \{\varphi(0) \in \linterpretations{\Sigma}{V} \with
      \varphi(r)\in X\)
      for some $r\in\reals_{\geq0}$ and (differentiable)
      \m{\varphi:[0,r]\to\linterpretations{\Sigma}{V}}
      such that
      \(\varphi(\zeta)\in\imodel{\I}{\ivr}\)
      and
      \m{\D[t]{\,\varphi(t)(x)} (\zeta) =
      \ivaluation{\iconcat[state=\varphi(\zeta)]{\I}}{f(x)}}
      for all \(0\leq\zeta\leq r\}\)
\item \(\strategyfor[\ptest{\ivr}]{X} = \imodel{\I}{\ivr}\cap X\)
\item \(\strategyfor[\pchoice{\alpha}{\beta}]{X} = \strategyfor[\alpha]{X}\cup\strategyfor[\beta]{X}\)
\item \(\strategyfor[\alpha;\beta]{X} = \strategyfor[\alpha]{\strategyfor[\beta]{X}}\)
\item \(\strategyfor[\prepeat{\alpha}]{X} = \capfold\{Z\subseteq\linterpretations{\Sigma}{V} \with X\cup\strategyfor[\alpha]{Z}\subseteq Z\}\)

\item \(\strategyfor[\pdual{\alpha}]{X} = \scomplement{(\strategyfor[\alpha]{\scomplement{X}})}\)
\end{enumerate}
The \emph{winning region} of Demon, i.e.\ the set of states \m{\dstrategyfor[\alpha]{X}} from which Demon has a winning strategy to achieve $X$ in $\alpha$ (whatever strategy Angel chooses) is defined inductively as follows
\begin{enumerate}
\item \(\dstrategyfor[\pupdate{\pumod{x}{\theta}}]{X} = \{\iportray{\I} \in \linterpretations{\Sigma}{V} \with \modif{\iget[state]{\I}}{x}{\ivaluation{\I}{\theta}} \in X\}\)
\item \(\dstrategyfor[\pevolvein{\D{x}=\genDE{x}}{\ivr}]{X} = \{\varphi(0) \in \linterpretations{\Sigma}{V} \with
      \varphi(r)\in X\)
      for all $r\in\reals_{\geq0}$ and (differentiable)
      \m{\varphi:[0,r]\to\linterpretations{\Sigma}{V}}
      such that
      \(\varphi(\zeta)\in\imodel{\I}{\ivr}\)
      and
      \m{\D[t]{\,\varphi(t)(x)} (\zeta) =
      \ivaluation{\iconcat[state=\varphi(\zeta)]{\I}}{\theta}}
      for all $0\leq\zeta\leq r\}$
\item \(\dstrategyfor[\ptest{\ivr}]{X} = \scomplement{(\imodel{\I}{\ivr})}\cup X\)
\item \(\dstrategyfor[\pchoice{\alpha}{\beta}]{X} = \dstrategyfor[\alpha]{X}\cap\dstrategyfor[\beta]{X}\)
\item \(\dstrategyfor[\alpha;\beta]{X} = \dstrategyfor[\alpha]{\dstrategyfor[\beta]{X}}\)
\item \(\dstrategyfor[\prepeat{\alpha}]{X} = \cupfold\{Z\subseteq\linterpretations{\Sigma}{V} \with Z\subseteq X\cap\dstrategyfor[\alpha]{Z}\}\)
\item \(\dstrategyfor[\pdual{\alpha}]{X} = \scomplement{(\dstrategyfor[\alpha]{\scomplement{X}})}\)
\end{enumerate}
\end{definition}
}

\providecommand{\axkey}[1]{\textcolor{vblue}{#1}}%
\cinferenceRuleStore[diamond|$\didia{\cdot}$]{diamond axiom}
{\linferenceRule[equiv]
  {\lnot\dbox{\ausprg}{\lnot \ausfml}}
  {\axkey{\ddiamond{\ausprg}{\ausfml}}}
}
{}
\cinferenceRuleStore[diamondax|$\didia{\cdot}$]{diamond axiom}
{\linferenceRule[equiv]
  {\lnot\dbox{\ausprgax}{\lnot \ausfmlax}}
  {\axkey{\ddiamond{\ausprgax}{\ausfmlax}}}
}
{}
\cinferenceRuleStore[assignb|$\dibox{:=}$]{assignment / substitution axiom}
{\linferenceRule[equiv]
  {p(\genDJ{x})}
  {\axkey{\dbox{\pupdate{\umod{x}{\genDJ{x}}}}{p(x)}}}
}
{}%
\cinferenceRuleStore[assignbax|$\dibox{:=}$]{assignment / substitution axiom}
{\linferenceRule[equiv]
  {p(\aconst)}
  {\axkey{\dbox{\pupdate{\umod{x}{\aconst}}}{p(x)}}}
}
{}%
\cinferenceRuleStore[Dassignb|$\dibox{:=}$]{differential assignment}
{\linferenceRule[equiv]
{p(\astrm)}
{\axkey{\dbox{\Dupdate{\Dumod{\D{x}}{\astrm}}}{p(\D{x})}}}
}
{}%
\cinferenceRuleStore[testb|$\dibox{?}$]{test}
{\linferenceRule[equiv]
  {(\ivr \limply \ausfml)}
  {\axkey{\dbox{\ptest{\ivr}}{\ausfml}}}
}{}%
\cinferenceRuleStore[testbax|$\dibox{?}$]{test}
{\linferenceRule[equiv]
  {(q \limply p)}
  {\axkey{\dbox{\ptest{q}}{p}}}
}{}%
\cinferenceRuleStore[evolveb|$\dibox{'}$]{evolve}
{\linferenceRule[equiv]
  {\lforall{t{\geq}0}{\dbox{\pupdate{\pumod{x}{\solf(t)}}}{p(x)}}}
  {\axkey{\dbox{\pevolve{\D{x}=\genDE{x}}}{p(x)}}}
}{\m{\D{\solf}(t)=\genDE{\solf}}}%
\cinferenceRuleStore[choiceb|$\dibox{\cup}$]{axiom of nondeterministic choice}
{\linferenceRule[equiv]
  {\dbox{\ausprg}{\ausfml} \land \dbox{\busprg}{\ausfml}}
  {\axkey{\dbox{\pchoice{\ausprg}{\busprg}}{\ausfml}}}
}{}%
\cinferenceRuleStore[choicebax|$\dibox{\cup}$]{axiom of nondeterministic choice}
{\linferenceRule[equiv]
  {\dbox{\ausprgax}{\ausfmlax} \land \dbox{\busprgax}{\ausfmlax}}
  {\axkey{\dbox{\pchoice{\ausprgax}{\busprgax}}{\ausfmlax}}}
}{}%
\cinferenceRuleStore[evolveinb|$\dibox{'}$]{evolve}
{\linferenceRule[equiv]
  {
        \lforall{t{\geq}0}{\big(
          (\lforall{0{\leq}s{\leq}t}{q(\solf(s))})
          \limply
          \dbox{\pupdate{\pumod{x}{\solf(t)}}}{p(x)}
        \big)}
      }
  {
        \dbox{\pevolvein{\D{x}=\genDE{x}}{q(x)}}{p(x)}
  }
}{}%
\cinferenceRuleStore[composeb|$\dibox{{;}}$]{composition} %
{\linferenceRule[equiv]
  {\dbox{\ausprg}{\dbox{\busprg}{\ausfml}}}
  {\axkey{\dbox{\ausprg;\busprg}{\ausfml}}}
}{}%
\cinferenceRuleStore[composebax|$\dibox{{;}}$]{composition} %
{\linferenceRule[equiv]
  {\dbox{\ausprgax}{\dbox{\busprgax}{\ausfmlax}}}
  {\axkey{\dbox{\ausprgax;\busprgax}{\ausfmlax}}}
}{}%
\cinferenceRuleStore[iterateb|$\dibox{{}^*}$]{iteration/repeat unwind} %
{\linferenceRule[equiv]
  {\ausfml \land \dbox{\ausprg}{\dbox{\prepeat{\ausprg}}{\ausfml}}}
  {\axkey{\dbox{\prepeat{\ausprg}}{\ausfml}}}
}{}%
\cinferenceRuleStore[iteratebax|$\dibox{{}^*}$]{iteration/repeat unwind} %
{\linferenceRule[equiv]
  {\ausfmlax \land \dbox{\ausprgax}{\dbox{\prepeat{\ausprgax}}{\ausfmlax}}}
  {\axkey{\dbox{\prepeat{\ausprgax}}{\ausfmlax}}}
}{}%
\cinferenceRuleStore[K|K]{K axiom / modal modus ponens}
{\linferenceRule[impl]
  {\dbox{\ausprg}{(\ausfml\limply\busfml)}}
  {(\dbox{\ausprg}{\ausfml}\limply\axkey{\dbox{\ausprg}{\busfml}})}
}{}%
\cinferenceRuleStore[Kax|K]{K axiom / modal modus ponens}
{\linferenceRule[impl]
  {\dbox{\ausprgax}{(\ausfmlax\limply\busfmlax)}}
  {(\dbox{\ausprgax}{\ausfmlax}\limply\axkey{\dbox{\ausprgax}{\busfmlax}})}
}{}%
\cinferenceRuleStore[I|II]{loop induction}
{\linferenceRule[impl]
  {\dbox{\prepeat{\ausprg}}{(\ausfml\limply\dbox{\ausprg}{\ausfml})}}
  {(\ausfml\limply\axkey{\dbox{\prepeat{\ausprg}}{\ausfml}})}
}{}%
\cinferenceRuleStore[Ieq|I]{loop induction}
{\linferenceRule[equiv]
  {\ausfml \land \dbox{\prepeat{\ausprg}}{(\ausfml\limply\dbox{\ausprg}{\ausfml})}}
  {\axkey{\dbox{\prepeat{\ausprg}}{\ausfml}}}
}{}%
\cinferenceRuleStore[Ieqax|I]{loop induction}
{\linferenceRule[equiv]
  {\ausfmlax \land \dbox{\prepeat{\ausprgax}}{(\ausfmlax\limply\dbox{\ausprgax}{\ausfmlax})}}
  {\axkey{\dbox{\prepeat{\ausprgax}}{\ausfmlax}}}
}{}%
\dinferenceRuleStore[backiterateb|\usebox{\backiterateb}]{backwards iteration/repeat unwind}
{\linferenceRule[equiv]
  {\ausfml \land \dbox{\prepeat{\ausprg}}{\dbox{\ausprg}{\ausfml}}}
  {\axkey{\dbox{\prepeat{\ausprg}}{\ausfml}}}
}{}%
\dinferenceRuleStore[iterateiterateb|$\dibox{{}^*{}^*}$]{double iteration}
{\linferenceRule[equiv]
  {\dbox{\prepeat{\ausprg}}{\ausfml}}
  {\axkey{\dbox{\prepeat{\ausprg};\prepeat{\ausprg}}{\ausfml}}}
}{}%
\dinferenceRuleStore[iterateiterated|$\didia{{}^*{}^*}$]{double iteration}
{\linferenceRule[equiv]
  {\ddiamond{\prepeat{\ausprg}}{\ausfml}}
  {\axkey{\ddiamond{\prepeat{\ausprg};\prepeat{\ausprg}}{\ausfml}}}
}{}%
\cinferenceRuleStore[B|B]{Barcan and converse}
{\linferenceRule[equiv]
        {\ddiamond{\ausprg}{\lexists{x}{\ausfml}}}
        {\lexists{x}{\ddiamond{\ausprg}{\ausfml}}}
}{\m{x{\not\in}\ausprg}}
\cinferenceRuleStore[V|V]{vacuous $\dbox{}{}$}
{\linferenceRule[impl]
  {p}
  {\axkey{\dbox{\ausprg}{p}}}
}{\m{FV(p)\cap BV(\ausprg)=\emptyset}}%
\cinferenceRuleStore[Vax|V]{vacuous $\dbox{}{}$}
{\linferenceRule[impl]
  {p}
  {\axkey{\dbox{a}{p}}}
}{}%
\cinferenceRuleStore[G|G]{$\dbox{}{}$ generalization} %
{\linferenceRule[formula]
  {\ausfml}
  {\dbox{\ausprg}{\ausfml}}
}{}%
\cinferenceRuleStore[Gax|G]{$\dbox{}{}$ generalization} %
{\linferenceRule[formula]
  {\ausfmlax}
  {\dbox{\ausprgax}{\ausfmlax}}
}{}%
\cinferenceRuleStore[genaax|$\forall{}$]{$\forall{}$ generalisation}
{\linferenceRule[formula]
  {p(x)}
  {\lforall{x}{p(x)}}
}{}%
\cinferenceRuleStore[MPax|MP]{modus ponens}
{\linferenceRule[formula]
  {p\limply q \quad p}
  {q}
}{}%
\cinferenceRuleStore[Mb|M${\dibox{\cdot}}$]{$\dbox{}{}$ monotone}
{\linferenceRule[formula]
  {\ausfml\limply \busfml}
  {\dbox{\ausprg}{\ausfml}\limply\dbox{\ausprg}{\busfml}}
}{}%
\cinferenceRuleStore[M|M]{$\ddiamond{}{}$ monotone / $\ddiamond{}{}$-generalization}
{\linferenceRule[formula]
  {\ausfml\limply\busfml}
  {\ddiamond{\ausprg}{\ausfml}\limply\ddiamond{\ausprg}{\busfml}}
}{}%

\dinferenceRuleStore[Mbr|M\rightrule]%
{$\ddiamond{}{}/\dbox{}{}$ generalization=M=G+K} 
{\linferenceRule[sequent]
  {\lsequent[L]{} {\dbox{\ausprg}{\busfml}} 
  &\lsequent[g]{\busfml} {\ausfml}}
  {\lsequent[L]{} {\dbox{\ausprg}{\ausfml}}}
}{}%

\dinferenceRuleStore[loop|loop]{inductive invariant}
{\linferenceRule[sequent]
  {\lsequent[L]{} {\inv}
  &\lsequent[g]{\inv} {\dbox{\ausprg}{\inv}}
  &\lsequent[g]{\inv} {\ausfml}}
  {\lsequent[L]{} {\dbox{\prepeat{\ausprg}}{\ausfml}}}
}{}%
\dinferenceRuleStore[invind|ind]{inductive invariant}
{\linferenceRule[sequent]
  {\lsequent[\globalrule]{\ausfml}{\dbox{\ausprg}{\ausfml}}}
  {\lsequent{\ausfml}{\dbox{\prepeat{\ausprg}}{\ausfml}}}
}{}%
\cinferenceRuleStore[con|con]{loop convergence right} 
{\linferenceRule[formula]
  {\lsequent[G]{\mapply{\var}{v}\land v>0}{\ddiamond{\ausprg}{\mapply{\var}{v-1}}}}
  {\lsequent[L]{\lexists{v}{\mapply{\var}{v}}}
      {\axkey{\ddiamond{\prepeat{\ausprg}}{\lexists{v{\leq}0}{\mapply{\var}{v}}}}}}
}{v\not\in\ausprg}%
\dinferenceRuleStore[congen|con]{loop convergence}
{\linferenceRule[sequent]
  {\lsequent[L]{}{\lexists{v}{\mapply{\var}{v}}}
  &\lsequent[G]{}{\lforall{v{>}0}{({\mapply{\var}{v}}\limply{\ddiamond{\ausprg}{\mapply{\var}{v-1}})}}}
  &\lsequent[G]{\lexists{v{\leq}0}{\mapply{\var}{v}}}{\busfml}
  }
  {\lsequent[L]{}{\ddiamond{\prepeat{\ausprg}}{\busfml}}}
}{v\not\in\ausprg}

\dinferenceRuleStore[band|${[]\land}$]{$\dbox{\cdot}{\land}$}
{\linferenceRule[equiv]
  {\dbox{\ausprg}{\ausfml} \land \dbox{\ausprg}{\busfml}}
  {\axkey{\dbox{\ausprg}{(\ausfml\land\busfml)}}}
}{}%

\dinferenceRuleStore[Hoarecompose|H${;}$]{Hoare $;$}
{\linferenceRule
  {A\limply\dbox{\ausprg}{E} & E\limply\dbox{\busprg}{B}}
  {A \limply \dbox{\ausprg;\busprg}{B}}
}{}%
\dinferenceRuleStore[composebrexplicit|$\dibox{{;}}$\rightrule]{$;$}
{\linferenceRule
  {A\limply\dbox{\ausprg}{\dbox{\busprg}{B}}}
  {A \limply \dbox{\ausprg;\busprg}{B}}
}{}
\cinferenceRuleStore[notr|$\lnot$\rightrule]{$\lnot$ right}
{\linferenceRule[sequent]
  {\lsequent[L]{\asfml}{}}
  {\lsequent[L]{}{\lnot \asfml}}
}{}%
\cinferenceRuleStore[notl|$\lnot$\leftrule]{$\lnot$ left}
{\linferenceRule[sequent]
  {\lsequent[L]{}{\asfml}}
  {\lsequent[L]{\lnot \asfml}{}}
}{}%
\cinferenceRuleStore[andr|$\land$\rightrule]{$\land$ right}
{\linferenceRule[sequent]
  {\lsequent[L]{}{\asfml}
    & \lsequent[L]{}{\bsfml}}
  {\lsequent[L]{}{\asfml \land \bsfml}}
}{}%
\cinferenceRuleStore[andl|$\land$\leftrule]{$\land$ left}
{\linferenceRule[sequent]
  {\lsequent[L]{\asfml , \bsfml}{}}
  {\lsequent[L]{\asfml \land \bsfml}{}}
}{}%
\cinferenceRuleStore[orr|$\lor$\rightrule]{$\lor$ right}
{\linferenceRule[sequent]
  {\lsequent[L]{}{\asfml, \bsfml}}
  {\lsequent[L]{}{\asfml \lor \bsfml}}
}{}%
\cinferenceRuleStore[orl|$\lor$\leftrule]{$\lor$ left}
{\linferenceRule[sequent]
  {\lsequent[L]{\asfml}{}
    & \lsequent[L]{\bsfml}{}}
  {\lsequent[L]{\asfml \lor \bsfml}{}}
}{}%
\cinferenceRuleStore[implyr|$\limply$\rightrule]{$\limply$ right}
{\linferenceRule[sequent]
  {\lsequent[L]{\asfml}{\bsfml}}
  {\lsequent[L]{}{\asfml \limply \bsfml}}
}{}%
\cinferenceRuleStore[implyl|$\limply$\leftrule]{$\limply$ left}
{\linferenceRule[sequent]
  {\lsequent[L]{}{\asfml}
    & \lsequent[L]{\bsfml}{}}
  {\lsequent[L]{\asfml \limply \bsfml}{}}
}{}%
\cinferenceRuleStore[equivr|$\lbisubjunct$\rightrule]{$\lbisubjunct$ right}
{\linferenceRule[sequent]
  {\lsequent[L]{\asfml}{\bsfml}
   & \lsequent[L]{\bsfml}{\asfml}}
  {\lsequent[L]{}{\asfml \lbisubjunct \bsfml}}
}{}%
\cinferenceRuleStore[equivl|$\lbisubjunct$\leftrule]{$\lbisubjunct$ left}
{\linferenceRule[sequent]
  {\lsequent[L]{\asfml\limply\bsfml, \bsfml\limply\asfml}{}}
  {\lsequent[L]{\asfml \lbisubjunct \bsfml}{}}
}{}%
\cinferenceRuleStore[id|id]{identity}
{\linferenceRule[sequent]
  {}
  {\lsequent[L]{\asfml}{\asfml}}
}{}%
\cinferenceRuleStore[cut|cut]{cut}
{\linferenceRule[sequent]
  {\lsequent[L]{}{\cusfml}
  &\lsequent[L]{\cusfml}{}}
  {\lsequent[L]{}{}}
}{}%
\cinferenceRuleStore[weakenr|W\rightrule]{weakening right}
{\linferenceRule[sequent]
  {\lsequent[L]{}{}}
  {\lsequent[L]{}{\asfml}}
}{}%
\cinferenceRuleStore[weakenl|W\leftrule]{weakening left}
{\linferenceRule[sequent]
  {\lsequent[L]{}{}}
  {\lsequent[L]{\asfml}{}}
}{}%
\cinferenceRuleStore[exchanger|P\rightrule]{exchange right}
{\linferenceRule[sequent]
  {\lsequent[L]{}{\bsfml,\asfml}}
  {\lsequent[L]{}{\asfml,\bsfml}}
}{}%
\cinferenceRuleStore[exchangel|P\leftrule]{exchange left}
{\linferenceRule[sequent]
  {\lsequent[L]{\bsfml,\asfml}{}}
  {\lsequent[L]{\asfml,\bsfml}{}}
}{}%
\cinferenceRuleStore[contractr|c\rightrule]{contract right}
{\linferenceRule[sequent]
  {\lsequent[L]{}{\asfml,\asfml}}
  {\lsequent[L]{}{\asfml}}
}{}%
\cinferenceRuleStore[contractl|c\leftrule]{contract left}
{\linferenceRule[sequent]
  {\lsequent[L]{\asfml,\asfml}{}}
  {\lsequent[L]{\asfml}{}}
}{}
\cinferenceRuleStore[closeTrue|$\top$\rightrule]{close by true}
{\linferenceRule[sequent]
  {}
  {\lsequent[L]{}{\ltrue}}
}{}%
\cinferenceRuleStore[closeFalse|$\bot$\leftrule]{close by false}
{\linferenceRule[sequent]
  {}
  {\lsequent[L]{\lfalse}{}}
}{}%

\cinferenceRuleStore[CE|CE]{congequiv congruence of equivalences on formulas}
{\linferenceRule[formula]
  {\ausfml \lbisubjunct \busfml}
  {\contextapp{C}{\ausfml} \lbisubjunct \contextapp{C}{\busfml}}
}{}%
\dinferenceRuleStore[CEr|CE\rightrule]{congequiv congruence of equivalences on formulas}
{\linferenceRule[formula]
  {\lsequent[L]{} {\contextapp{C}{\busfml}}
  &\lsequent[g]{} {\ausfml \lbisubjunct \busfml}}
  {\lsequent[L]{} {\contextapp{C}{\ausfml}}}
}{}%
\dinferenceRuleStore[CEl|CE\leftrule]{congequiv congruence of equivalences on formulas}
{\linferenceRule[formula]
  {\lsequent[L]{\contextapp{C}{\busfml}} {}
  &\lsequent[g]{} {\ausfml \lbisubjunct \busfml}}
  {\lsequent[L]{\contextapp{C}{\ausfml}} {}}
}{}%

\cinferenceRuleStore[allr|$\forall$\rightrule]{$\lforall{}{}$ right}
{\linferenceRule[sequent]
  {\lsequent[L]{}{p(y)}}
  {\lsequent[L]{}{\lforall{x}{p(x)}}}
}{\m{y\not\in\Gamma{,}\Delta{,}\lforall{x}{p(x)}}}%
\cinferenceRuleStore[alll|$\forall$\leftrule]{$\lforall{}{}$ left instantiation}
{\linferenceRule[sequent]
  {\lsequent[L]{p(\astrm)}{}}
  {\lsequent[L]{\lforall{x}{p(x)}}{}}
}{arbitrary term $\astrm$}%
\cinferenceRuleStore[existsr|$\exists$\rightrule]{$\lexists{}{}$ right}
{\linferenceRule[sequent]
  {\lsequent[L]{}{p(\astrm)}}
  {\lsequent[L]{}{\lexists{x}{p(x)}}}
}{arbitrary term $\astrm$}%
\cinferenceRuleStore[existsl|$\exists$\leftrule]{$\lexists{}{}$ left}
{\linferenceRule[sequent]
  {\lsequent[L]{p(y)} {}}
  {\lsequent[L]{\lexists{x}{p(x)}} {}}
}{\m{y\not\in\Gamma{,}\Delta{,}\lexists{x}{p(x)}}}%

\cinferenceRuleStore[qear|\usebox{\Rval}]{quantifier elimination real arithmetic}
{\linferenceRule[sequent]
  {}
  {\lsequent[g]{\Gamma}{\Delta}}
}{\text{if}~\landfold_{\ausfml\in\Gamma} \ausfml \limply \lorfold_{\busfml\in\Delta} \busfml ~\text{is valid in \LOS[\reals]}}%

\dinferenceRuleStore[allGi|i$\forall$]{inverse universal generalization / universal instantiation}
{\linferenceRule[sequent]
  {\lsequent[L]{} {\lforall{x}{\ausfml}}}
  {\lsequent[L]{} {\ausfml}}
}{}%

\cinferenceRuleStore[applyeqr|=\rightrule]{apply equation}
{\linferenceRule[sequent]
  {\lsequent[L]{x=\astrm}{p(\astrm)}}
  {\lsequent[L]{x=\astrm}{p(x)}}
}{}%
\cinferenceRuleStore[applyeql|=\leftrule]{apply equation}
{\linferenceRule[sequent]
  {\lsequent[L]{x=\astrm,p(\astrm)}{}}
  {\lsequent[L]{x=\astrm,p(x)}{}}
}{}%

\dinferenceRuleStore[alldupl|$\forall\forall$\leftrule]{$\lforall{}{}$ left instantiation retaining duplicates}
{\linferenceRule[sequent]
  {\lsequent[L]{\lforall{x}{p(x)},p(\astrm)}{}}
  {\lsequent[L]{\lforall{x}{p(x)}}{}}
}{}%

\dinferenceRuleStore[choicebrinsist|$\dibox{\cup}\rightrule$]{}
{\linferenceRule
  {\lsequent[L]{}{\dbox{\asprg}{\ausfml}\land\dbox{\bsprg}{\ausfml}}}
  {\lsequent[L]{}{\dbox{\pchoice{\asprg}{\bsprg}}{\ausfml}}}
}{}
\dinferenceRuleStore[choiceblinsist|$\dibox{\cup}\leftrule$]{}
{\linferenceRule
  {\lsequent[L]{\dbox{\asprg}{\ausfml}\land\dbox{\bsprg}{\ausfml}}{}}
  {\lsequent[L]{\dbox{\pchoice{\asprg}{\bsprg}}{\ausfml}}{}}
}{}
\dinferenceRuleStore[choicebrinsist2|$\dibox{\cup}\rightrule2$]{}
{\linferenceRule
  {\lsequent[L]{}{\dbox{\asprg}{\ausfml}}
  &\lsequent[L]{}{\dbox{\bsprg}{\ausfml}}}
  {\lsequent[L]{}{\dbox{\pchoice{\asprg}{\bsprg}}{\ausfml}}}
}{}
\dinferenceRuleStore[choiceblinsist2|$\dibox{\cup}\leftrule2$]{}
{\linferenceRule
  {\lsequent[L]{\dbox{\asprg}{\ausfml},\dbox{\bsprg}{\ausfml}}{}}
  {\lsequent[L]{\dbox{\pchoice{\asprg}{\bsprg}}{\ausfml}}{}}
}{}
\dinferenceRuleStore[cutr|cut\rightrule]{cut right}
{\linferenceRule[sequent]
  {\lsequent[L]{}{\bsfml}
  &\lsequent[L]{}{\bsfml\limply\asfml}}
  {\lsequent[L]{}{\asfml}}
}{}
\dinferenceRuleStore[cutl|cut\leftrule]{cut left}
{\linferenceRule[sequent]
  {\lsequent[L]{\bsfml} {}
  &\lsequent[L]{}{\asfml\limply\bsfml}}
  {\lsequent[L]{\asfml} {}}
}{}

\cinferenceRuleStore[Dplus|$+'$]{derive sum}
{\linferenceRule[eq]
  {\der{\asdtrm}+\der{\bsdtrm}}
  {\axkey{\der{\asdtrm+\bsdtrm}}}
}
{}
\cinferenceRuleStore[Dplusax|$+'$]{derive sum}
{\linferenceRule[eq]
  {\der{\asdtrmax}+\der{\bsdtrmax}}
  {\axkey{\der{\asdtrmax+\bsdtrmax}}}
}
{}
\cinferenceRuleStore[Dminus|$-'$]{derive minus}
{\linferenceRule[eq]
  {\der{\asdtrm}-\der{\bsdtrm}}
  {\axkey{\der{\asdtrm-\bsdtrm}}}
}
{}
\cinferenceRuleStore[Dminusax|$-'$]{derive minus}
{\linferenceRule[eq]
  {\der{\asdtrmax}-\der{\bsdtrmax}}
  {\axkey{\der{\asdtrmax-\bsdtrmax}}}
}
{}
\cinferenceRuleStore[Dtimes|$\cdot'$]{derive product}
{\linferenceRule[eq]
  {\der{\asdtrm}\cdot \bsdtrm+\asdtrm\cdot\der{\bsdtrm}}
  {\axkey{\der{\asdtrm\cdot \bsdtrm}}}
}
{}
\cinferenceRuleStore[Dtimesax|$\cdot'$]{derive product}
{\linferenceRule[eq]
  {\der{\asdtrmax}\cdot \bsdtrmax+\asdtrmax\cdot\der{\bsdtrmax}}
  {\axkey{\der{\asdtrmax\cdot \bsdtrmax}}}
}
{}
\cinferenceRuleStore[Dquotient|$/'$]{derive quotient}
{\linferenceRule[eq]
  {\big(\der{\asdtrm}\cdot \bsdtrm-\asdtrm\cdot\der{\bsdtrm}\big) / \bsdtrm^2}
  {\axkey{\der{\asdtrm/\bsdtrm}}}
}
{}
\cinferenceRuleStore[Dquotientax|$/'$]{derive quotient}
{\linferenceRule[eq]
  {\big(\der{\asdtrmax}\cdot \bsdtrmax-\asdtrmax\cdot\der{\bsdtrmax}\big) / \bsdtrmax^2}
  {\axkey{\der{\asdtrmax/\bsdtrmax}}}
}
{}
\cinferenceRuleStore[Dconst|$c'$]{derive constant}
{\linferenceRule[eq]
  {0}
  {\axkey{\der{\aconst}}}
  \hspace{3cm}
}
{\text{for numbers or constants~$\aconst$}}%
\cinferenceRuleStore[Dvar|$x'$]{derive variable}
{\linferenceRule[eq]
  {\D{x}}
  {\axkey{\der{x}}}
}
{\text{for variable~$x\in\allvars$}}%

\cinferenceRuleStore[DE|DE]{differential effect} %
{\linferenceRule[viuqe]
  {\axkey{\dbox{\pevolvein{\D{x}=\genDE{x}}{\ivr}}{\ausfml}}}
  {\dbox{\pevolvein{\D{x}=\genDE{x}}{\ivr}}{\dbox{\axeffect{\Dupdate{\Dumod{\D{x}}{\genDE{x}}}}}{\ausfml}}}
}
{}%
\cinferenceRuleStore[DEax|DE]{differential effect} %
{\linferenceRule[viuqe]
  {\axkey{\dbox{\pevolvein{\D{x}=\genDE{x}}{q(x)}}{\ausfmlax}}}
  {\dbox{\pevolvein{\D{x}=\genDE{x}}{q(x)}}{\dbox{\axeffect{\Dupdate{\Dumod{\D{x}}{\genDE{x}}}}}{\ausfmlax}}}
}
{}%

\cinferenceRuleStore[Dand|${\land}'$]{derive and}
{\linferenceRule[equiv]
  {\der{\asfml}\land\der{\bsfml}}
  {\axkey{\der{\asfml\land\bsfml}}}
}
{}
\cinferenceRuleStore[Dor|${\lor}'$]{derive or}
{\linferenceRule[equiv]
  {\der{\asfml}\land\der{\bsfml}}
  {\axkey{\der{\asfml\lor\bsfml}}}
}
{}
\cinferenceRuleStore[diffweaken|DW]{differential evolution domain} %
{\linferenceRule[viuqe]
  {\axkey{\dbox{\pevolvein{\D{x}=\genDE{x}}{\ivr}}{\ousfml[x]}}}
  {\dbox{\pevolvein{\D{x}=\genDE{x}}{\ivr}}{(\axeffect{\ivr}\limply \ousfml[x])}}
}
{}%
\cinferenceRuleStore[diffweakenax|DW]{differential evolution domain} 
{\linferenceRule[viuqe]
  {\axkey{\dbox{\pevolvein{\D{x}=\genDE{x}}{q(x)}}{p(x)}}}
  {\dbox{\pevolvein{\D{x}=\genDE{x}}{q(x)}}{(\axeffect{q(x)}\limply p(x))}}
}
{}%
\cinferenceRuleStore[dW|dW]{differential weakening}
{\linferenceRule[sequent]
  {\lsequent[g]{\ivr} {\ousfml[x]}}
  {\lsequent[g]{\Gamma} {\dbox{\pevolvein{\D{x}=f(x)}{\ivr}}{\ousfml[x]},\Delta}}
}
{}%
\cinferenceRuleStore[DI|DI]{differential induction}
{\linferenceRule[lpmi]
  {\big(\axkey{\dbox{\pevolvein{\D{x}=\genDE{x}}{\ivr}}{\ousfml[x]}}
  \lbisubjunct \dbox{\ptest{\ivr}}{\ousfml[x]}\big)}
  {(\ivr\limply\dbox{\pevolvein{\D{x}=\genDE{x}}{\ivr}}{\axeffect{\der{\ousfml[x]}}})}
}
{}%
\cinferenceRuleStore[DIax|DI]{differential induction}
{\linferenceRule[lpmi]
  {\big(\axkey{\dbox{\pevolvein{\D{x}=\genDE{x}}{q(x)}}{p(x)}}
  \lbisubjunct \dbox{\ptest{q(x)}}{p(x)}\big)}
  {(q(x)\limply\dbox{\pevolvein{\D{x}=\genDE{x}}{q(x)}}{\axeffect{\der{p(x)}}})}
}
{}%
\cinferenceRuleStore[DIlight|DI]{differential induction}
{\linferenceRule[lpmi]
  {\big(\axkey{\dbox{\pevolvein{\D{x}=\genDE{x}}{\ivr}}{\ousfml[x]}}
  \lbisubjunct \dbox{\ptest{\ivr}}{\ousfml[x]}\big)}
  {\dbox{\pevolvein{\D{x}=\genDE{x}}{\ivr})}{\axeffect{\der{\ousfml[x]}}}}
}
{}%

\cinferenceRuleStore[dI|dI]{differential invariant}
{\linferenceRule[sequent]
  {\lsequent[g]{\ivr}{\Dusubst{\D{x}}{\genDE{x}}{\der{F}}}}
  {\lsequent{F}{\dbox{\pevolvein{\D{x}=\genDE{x}}{\ivr}}{F}}}
}{}
\cinferenceRuleStore[DC|DC]{differential cut}
{\linferenceRule[lpmi]
  {\big(\axkey{\dbox{\pevolvein{\D{x}=\genDE{x}}{\ivr}}{\ousfml[x]}} \lbisubjunct \dbox{\pevolvein{\D{x}=\genDE{x}}{\ivr\land \axeffect{\ousfmlc[x]}}}{\ousfml[x]}\big)}
  {\dbox{\pevolvein{\D{x}=\genDE{x}}{\ivr}}{\axeffect{\ousfmlc[x]}}}
}
{}%
\cinferenceRuleStore[DCax|DC]{differential cut}
{\linferenceRule[lpmi]
  {\big(\axkey{\dbox{\pevolvein{\D{x}=\genDE{x}}{q(x)}}{p(x)}} \lbisubjunct \dbox{\pevolvein{\D{x}=\genDE{x}}{q(x)\land \axeffect{r(x)}}}{p(x)}\big)}
  {\dbox{\pevolvein{\D{x}=\genDE{x}}{q(x)}}{\axeffect{r(x)}}}
}
{}%
\cinferenceRuleStore[dC|dC]{differential cut}%
{\linferenceRule[sequent]
  {\lsequent[L]{}{\dbox{\pevolvein{\D{x}=\genDE{x}}{\ivr}}{\axeffect{\cusfml}}}
  &\lsequent[L]{}{\dbox{\pevolvein{\D{x}=\genDE{x}}{(\ivr\land \axeffect{\cusfml})}}{\ousfml[x]}}}
  {\lsequent[L]{}{\dbox{\pevolvein{\D{x}=\genDE{x}}{\ivr}}{\ousfml[x]}}}
}{}
\cinferenceRuleStore[DGanyode|DG]{differential ghost variables (unsound!)}
{\linferenceRule[viuqe]
  {\axkey{\dbox{\pevolvein{\D{x}=\genDE{x}}{\ivr}}{\ousfml[x]}}}
  {\lexists{y}{\dbox{\pevolvein{\D{x}=\genDE{x}\syssep\axeffect{\D{y}=g(x,y)}}{\ivr}}{\ousfml[x]}}}
}
{}
\cinferenceRuleStore[DG|DG]{differential ghost variables}
{\linferenceRule[viuqe]
  {\axkey{\dbox{\pevolvein{\D{x}=\genDE{x}}{\ivr}}{\ousfml[x]}}}
  {\lexists{y}{\dbox{\pevolvein{\D{x}=\genDE{x}\syssep\axeffect{\D{y}=a(x)\cdot y+b(x)}}{\ivr}}{\ousfml[x]}}}
}
{}
\cinferenceRuleStore[DGax|DG]{differential ghost variables}
{\linferenceRule[viuqe]
  {\axkey{\dbox{\pevolvein{\D{x}=\genDE{x}}{q(x)}}{p(x)}}}
  {\lexists{y}{\dbox{\pevolvein{\D{x}=\genDE{x}\syssep\axeffect{\D{y}=a(x)\cdot y+b(x)}}{q(x)}}{p(x)}}}
}
{}
\cinferenceRuleStore[dG|dG]{dG}%
{\linferenceRule[sequent]
  {\lsequent[L]{} {\lexists{y}{\dbox{\pevolvein{\D{x}=f(x)\syssep\axeffect{\D{y}=a(x)\cdot y+b(x)}}{\oivr[x]}}{\ousfml[x]}}}
  }
  {\lsequent[L]{} {\dbox{\pevolvein{\D{x}=f(x)}{\oivr[x]}}{\ousfml[x]}}}
}
{}%

\dinferenceRuleStore[assignbeqr|$\dibox{:=}_=$]{assignb}%
  {\linferenceRule[sequent]
    {\lsequent[L]{y=\austrm} {p(y)}}
    {\lsequent[L]{} {\dbox{\pupdate{\umod{x}{\austrm}}}{p(x)}}}
  }
  {\text{$y$ new}}

\cinferenceRuleStore[DSax|DS]{(constant) differential equation solution} 
{\linferenceRule[viuqe]
  {\axkey{\dbox{\pevolvein{\D{x}=\aconst}{q(x)}}{p(x)}}}
  {\lforall{t{\geq}0}{\big((\lforall{0{\leq}s{\leq}t}{q(x+\aconst\itimes s)}) \limply \dbox{\pupdate{\pumod{x}{x+\aconst\itimes t}}}{p(x)}\big)}}
}
{}

\dinferenceRuleStore[DIeq0|DI]{differential invariant axiom}
{\linferenceRule[lpmi]
  {\big(\axkey{\dbox{\pevolve{\D{x}=\genDE{x}}}{\,\astrm=0}} \lbisubjunct \astrm=0\big)}
  {\dbox{\pevolve{\D{x}=\genDE{x}}}{\,\axeffect{\der{\astrm}=0}}}
}
{}
\dinferenceRuleStore[diffindeq0|dI]{differential invariant $=0$ case}
{\linferenceRule[sequent]
  {\lsequent{~}{\Dusubst{\D{x}}{\genDE{x}}{\der{\astrm}}=0}}
  {\lsequent{\astrm=0}{\dbox{\pevolve{\D{x}=\genDE{x}}}{\astrm=0}}}
}{}
\cinferenceRuleStore[Liec|dI$_c$]{}
{\linferenceRule
  {\lsequent{\ivr}{\Dusubst{\D{x}}{\genDE{x}}{\der{\astrm}}=0}}
  {\lsequent{}{\lforall{c}{\big(\astrm=c \limply \dbox{\pevolvein{\D{x}=\genDE{x}}{\ivr}}{\astrm=c}\big)}}}
}{}

\cinferenceRuleStore[DIeq|DI$_=$]{differential induction $=$ case}
{\linferenceRule[lpmi]
  {\big(\axkey{\dbox{\pevolvein{\D{x}=\genDE{x}}{\ivr}}{\asdtrm=\bsdtrm}}
  \lbisubjunct \dbox{\ptest{\ivr}}{\asdtrm=\bsdtrm}\big)}
  {\dbox{\pevolvein{\D{x}=\genDE{x}}{\ivr})}{\axeffect{\der{\asdtrm}=\der{\bsdtrm}}}}
}
{}%

\dinferenceRuleStore[diffindgen|dI']{differential invariant}
{\linferenceRule[sequent]
  {\lsequent[L]{}{\inv}
  &\lsequent[g]{\ivr}{\Dusubst{\D{x}}{\genDE{x}}{\der{\inv}}}
  &\lsequent[g]{\inv}{\psi}
  }
  {\lsequent[L]{}{\dbox{\pevolvein{\D{x}=\genDE{x}}{\ivr}}{\psi}}}
}{}
\cinferenceRuleStore[diffindunsound|dI$_{??}$]{unsound}
{\linferenceRule[sequent]
  {\lsequent{\ivr\land\inv}{\Dusubst{\D{x}}{\genDE{x}}{\der{\inv}}}}
  {\lsequent{\inv}{\dbox{\pevolvein{\D{x}=\genDE{x}}{\ivr}}{\inv}}}
}{}

\dinferenceRuleStore[introaux|iG]{introduce discrete ghost variable}
{\linferenceRule[sequent]
  {\lsequent[L]{}{\dbox{\axeffect{\pupdate{\pumod{y}{\astrm}}}}{p}}}
  {\lsequent[L]{} {p}}
}{\text{$y$ new}}%
\dinferenceRuleStore[diffaux|dA]{differential auxiliary variables}
{\linferenceRule[sequent]
  {\lsequent[\globalrule]{}{\inv\lbisubjunct\lexists{y}{G}}
  &\lsequent{G} {\dbox{\pevolvein{\D{x}=\genDE{x}\syssep\axeffect{\D{y}=a(x)\cdot y+b(x)}}{\ivr}}{G}}}
  {\lsequent{\inv} {\dbox{\pevolvein{\D{x}=\genDE{x}}{\ivr}}{\inv}}}
}{}%
\cinferenceRuleStore[randomd|$\didia{{:}*}$]{nondeterministic assignment}
{\linferenceRule[equiv]
  {\lexists{x}{\ousfml[x]}}
  {\axkey{\ddiamond{\prandom{x}}{\ousfml[x]}}}
}{}
\cinferenceRuleStore[randomb|$\dibox{{:}*}$]{nondeterministic assignment}
{\linferenceRule[equiv]
  {\lforall{x}{\ousfml[x]}}
  {\axkey{\dbox{\prandom{x}}{\ousfml[x]}}}
}{}

\cinferenceRuleStore[box|$\dibox{\cdot}$]{box axiom}
{\linferenceRule[equiv]
  {\lnot\ddiamond{\ausprg}{\lnot\ausfml}}
  {\axkey{\dbox{\ausprg}{\ausfml}}}
}
{}
\cinferenceRuleStore[assignd|$\didia{:=}$]{assignment / substitution axiom}
{\linferenceRule[equiv]
  {p(\genDJ{x})}
  {\axkey{\ddiamond{\pupdate{\umod{x}{\genDJ{x}}}}{p(x)}}}
}
{}%
\cinferenceRuleStore[evolved|$\didia{'}$]{evolve}
{\linferenceRule[equiv]
  {\lexists{t{\geq}0}{\ddiamond{\pupdate{\pumod{x}{\solf(t)}}}{p(x)}}\hspace{1cm}}
  {\axkey{\ddiamond{\pevolve{\D{x}=\genDE{x}}}{p(x)}}}
}{\m{\D{\solf}(t)=\genDE{\solf}}}%
\cinferenceRuleStore[evolveind|$\didia{'}$]{evolve}
{\linferenceRule[equiv]
  {\lexists{t{\geq}0}{\big((\lforall{0{\leq}s{\leq}t}{q(\solf(s))}) \land 
  \ddiamond{\pupdate{\pumod{x}{\solf(t)}}}{p(x)}\big)}}
  {\axkey{\ddiamond{\pevolvein{\D{x}=\genDE{x}}{q(x)}}{p(x)}}}
}{\m{\D{\solf}(t)=\genDE{\solf}}}%
\cinferenceRuleStore[testd|$\didia{?}$]{test}
{\linferenceRule[equiv]
  {\ivr \land \ausfml}
  {\axkey{\ddiamond{\ptest{\ivr}}{\ausfml}}}
}{}
\cinferenceRuleStore[choiced|$\didia{\cup}$]{axiom of nondeterministic choice}
{\linferenceRule[equiv]
  {\ddiamond{\ausprg}{\ausfml} \lor \ddiamond{\busprg}{\ausfml}}
  {\axkey{\ddiamond{\pchoice{\ausprg}{\busprg}}{\ausfml}}}
}{}
\cinferenceRuleStore[composed|$\didia{{;}}$]{composition}
{\linferenceRule[equiv]
  {\ddiamond{\ausprg}{\ddiamond{\busprg}{\ausfml}}}
  {\axkey{\ddiamond{\ausprg;\busprg}{\ausfml}}}
}{}
\cinferenceRuleStore[iterated|$\didia{{}^*}$]{iteration/repeat unwind pre-fixpoint, even fixpoint}
{\linferenceRule[equiv]
  {\ausfml \lor \ddiamond{\ausprg}{\ddiamond{\prepeat{\ausprg}}{\ausfml}}}
  {\axkey{\ddiamond{\prepeat{\ausprg}}{\ausfml}}}
}{}
\cinferenceRuleStore[duald|$\didia{{^d}}$]{dual}
{\linferenceRule[equiv]
  {\lnot\ddiamond{\ausprg}{\lnot\ausfml}}
  {\axkey{\ddiamond{\pdual{\ausprg}}{\ausfml}}}
}{}
\cinferenceRuleStore[dualb|$\dibox{{^d}}$]{dual}
{\linferenceRule[equiv]
  {\lnot\dbox{\ausprg}{\lnot\ausfml}}
  {\axkey{\dbox{\pdual{\ausprg}}{\ausfml}}}
}{}
\cinferenceRuleStore[FP|FP]{iteration is least fixpoint / reflexive transitive closure RTC, equivalent to invind in the presence of M}
{\linferenceRule[formula]
  {\ausfml \lor \ddiamond{\ausprg}{\busfml} \limply \busfml}
  {\ddiamond{\prepeat{\ausprg}}{\ausfml} \limply \busfml}
}{}
\cinferenceRuleStore[invindg|ind]{inductive invariant for games}
{\linferenceRule[formula]
  {\ausfml\limply\dbox{\ausprg}{\ausfml}}
  {\ausfml\limply\dbox{\prepeat{\ausprg}}{\ausfml}}
}{}%

\dinferenceRuleStore[dchoiced|$\didia{{\cap}}$]{Demon's choice}
{
\axkey{\ddiamond{\dchoice{\ausprg}{\busprg}}{\ausfml}} \lbisubjunct \ddiamond{\ausprg}{\ausfml} \land \ddiamond{\busprg}{\ausfml}
}{}
\dinferenceRuleStore[dchoiceb|$\dibox{{\cap}}$]{Demon's choice}
{
\axkey{\dbox{\dchoice{\ausprg}{\busprg}}{\ausfml}} \lbisubjunct \dbox{\ausprg}{\ausfml} \lor \dbox{\busprg}{\ausfml}
}{}
\dinferenceRuleStore[diterateb|$\dibox{\drepeat{}}$]{Demon's repetition}
{\linferenceRule[equiv]
  {\ausfml \lor \dbox{\ausprg}{\dbox{\drepeat{\ausprg}}{\ausfml}}}
  {\axkey{\dbox{\drepeat{\ausprg}}{\ausfml}}}
}{}
\dinferenceRuleStore[diterated|$\didia{\drepeat{}}$]{Demon's repetition}
{\linferenceRule[equiv]
  {\ausfml \land \ddiamond{\ausprg}{\ddiamond{\drepeat{\ausprg}}}{\ausfml}}
  {\axkey{\ddiamond{\drepeat{\ausprg}}{\ausfml}}}
}{}
\dinferenceRuleStore[dinvindg|ind$\drepeat{}$]{inductive invariant for games}
{\linferenceRule[formula]
  {\ausfml\limply\ddiamond{\ausprg}{\ausfml}}
  {\ausfml\limply\ddiamond{\drepeat{\ausprg}}{\ausfml}}
}{}
\dinferenceRuleStore[dFP|FP$\drepeat{}$]{dual iteration is least fixpoint in Demon's winning strategy}
{\linferenceRule[formula]
  {\ausfml \lor \dbox{\ausprg}{\busfml} \limply \busfml}
  {\dbox{\drepeat{\ausprg}}{\ausfml} \limply \busfml}
}{}

\cinferenceRuleStore[US|US]{uniform substitution}
{\linferenceRule[formula]
  {\phi}
  {\applyusubst{\sigma}{\phi}}
}{}%

\cinferenceRuleStore[linequs|$\exists$lin]{linear equation uniform substitution}
{\linferenceRule[impl]
  {b\neq0}
  {\big(\lexists{x}{(b\cdot x+c=0 \land q(x))}
  \lbisubjunct {q(-c/b)}\big)}
}{}

\dinferenceRuleStore[FA|FA]{First arrival}
{\ddiamond{\prepeat{\ausprg}}{\ausfml} \limply \ausfml \lor \ddiamond{\prepeat{\ausprg}}{(\lnot\ausfml\land\ddiamond{\ausprg}{\ausfml})}
}{}
\dinferenceRuleStore[Mor|M]{monotonicity axiom}
{\ddiamond{\ausprg}{(\ausfml\lor\busfml)}
\lbisubjunct
\ddiamond{\ausprg}{\ausfml} \lor \ddiamond{\ausprg}{\busfml}
}{}
\dinferenceRuleStore[VK|VK]{vacuous possible $\dbox{}{}$}
{\linferenceRule[impl]
  {p}
  {(\dbox{\ausprg}{\ltrue}{\limply}\dbox{\ausprg}{p})}
  \qquad
}{\m{\freevars{p}\cap \boundvars{\ausprg}=\emptyset}}
\dinferenceRuleStore[R|R]{Regular}
{\linferenceRule[formula]
  {\ausfml_1\land\ausfml_2\limply\busfml}
  {\dbox{\ausprg}{\ausfml_1} \land \dbox{\ausprg}{\ausfml_2} \limply \dbox{\ausprg}{\busfml}}
}{}

\newcommand{\solf}{y}
\begin{figure}[tbhp]
  \centering
  \begin{calculuscollections}{\columnwidth}
    \begin{calculus}
      \cinferenceRuleQuote{box}
      \cinferenceRuleQuote{assignd}
      \cinferenceRuleQuote{evolved}
      \cinferenceRuleQuote{testd}
      \cinferenceRuleQuote{choiced}
      \cinferenceRuleQuote{composed}
      \cinferenceRuleQuote{iterated}
      \cinferenceRuleQuote{duald}
      \cinferenceRuleQuote{dchoiced}
      \cinferenceRuleQuote{diterated}
      \cinferenceRuleQuote{assignb}
      \cinferenceRuleQuote{evolveb}
      \cinferenceRuleQuote{testb}
      \cinferenceRuleQuote{choiceb}
      \cinferenceRuleQuote{composeb}
      \cinferenceRuleQuote{iterateb}
      \cinferenceRuleQuote{dualb}
      \cinferenceRuleQuote{dchoiceb}
      \cinferenceRuleQuote{diterateb}
    \end{calculus}

    \begin{calculus}
      \cinferenceRuleQuote{loop}
      \cinferenceRuleQuote{M}
      \cinferenceRuleQuote{Mb}
    \end{calculus}
    \begin{calculus}
      \dinferenceRuleQuote{invindg}
      \cinferenceRuleQuote{FP}
      \cinferenceRuleQuote{dFP}
    \end{calculus}
  \end{calculuscollections}
  \caption{\dGL axiomatization and derived axioms and rules}
  \label{fig:dGL-calculus}
\end{figure}

\subsection{Lemmas and Theorems from the Main Text}

\printProofs

\section{Parachute: \reduce of Non-solvable Dynamics}\label{app:parachute-ode}

The parachute benchmark (\rref{app:parachute}) has non-solvable dynamics whose exact solution involves hyperbolic functions.
The differential equations are $\{x' = -v, v' = -r v^2 + g\}$ and admit the solution: \[\left\{v(t)\to \frac{\sqrt{g} \tanh \left(\sqrt{g} \sqrt{r} t+c_1 \sqrt{g} \sqrt{r}\right)}{\sqrt{r}},x(t)\to c_2-\frac{\log \left(\cosh \left(\sqrt{g} \sqrt{r} (t+c_1)\right)\right)}{r}\right\}.\]
This solution does not belong to the decidable fragment of arithmetic, and QE is not guaranteed to terminate.
Reduce is able to still return preconditions over this ODE by using differential invariants generated by Pegasus.

The first call to reduce is for program $[\{x'=-v, v'=-rv^2+g, t'=1 \,\&\, v>0 \land t \le T\}](x<0 \lor v<m)$.
A call to Pegasus identifies the first degree Darboux polynomials of this ODE to be $\left\{\sqrt{g},\sqrt{g}-\sqrt{r} v,\sqrt{g}+\sqrt{r} v\right\}$.
As a heuristic to retain only the most useful polynomials, we filter out terms that are purely constant.
The Darboux rule of \dL tells us that the following formulas are now invariants of the differential equation system: $\left\{\sqrt{g}-\sqrt{r} v\geq 0,\sqrt{g}+\sqrt{r} v \geq 0 \right\}$.

When do these invariants imply the desired post condition, $x<0 \lor v<m$?
Mathematically, the answer is the formula
\[ \sqrt{g}-\sqrt{r} v\geq 0\land \sqrt{g}+\sqrt{r} v\geq 0 \limply x<0\lor v<m.\]
To find an invariant that implies this formula, we eliminate variables $x$ and $v$ using quantifier elimination.
The resulting expression is $m>0\land g\geq 0\land g<m^2 r$.
Thus, \reduce returns $\sqrt{g}-\sqrt{r} v\geq 0\land \sqrt{g}+\sqrt{r} v\geq 0 \land m>0\land g\geq 0\land g<m^2 r$ as a precondition.
This expression is invariant under the problem's dynamics and implies the postcondition $x<0\lor v<m$.

\section{Benchmarks}\label{app:benchmarks}

This section lists all the benchmarks proposed.
It also shows the solution of each benchmark, annotated with the meaning of the solution expressions.
The algebraic formulas presented are synthesized by CESAR automatically.
The annotations are added manually for the convenience of the reader.
\rref{tab:benchmarks} provides a one-sentence summary for each benchmark problem.

\begin{table}[tbhp]
  \caption{Benchmark listing.}
  \label{tab:benchmarks}
  \centering
  \begin{tabularx}{\textwidth}{lX}
    \toprule
    \textbf{Benchmark} & \textbf{Description} \\
    \midrule
    ETCS Train & Kernel of the European Train Control System case study \cite{DBLP:conf/icfem/PlatzerQ09}. \\
    Sled         & Swerve to avoid a wall. \\
    Intersection & Car must either cross an intersection before the light turns red, or stop before the intersection. It may never stop at the intersection. \\
    Curvebot     & A Dubin's car must avoid an obstacle. \\
    Parachute    & Use the irreversible choice to open a parachute with some air resistance to stay at a safe velocity. Drawn form \cite{DBLP:conf/itp/FultonMBP17}.\\
    Corridor     & Navigate a corridor with a side passage and dead end. \\
    Power Station & Choose between producing and distributing power while dealing with resistance loss and trying to meet a quota. \\
    Coolant       & A coolant system in an energy plant must maintain sufficient heat absorption while meeting coolant discharge limits. \\
    \bottomrule
  \end{tabularx}
  \end{table}

\subsection{ETCS Train}
\label{app:etcs}

The European Train Control System has been systematically but manually modeled and proved safe in the literature \cite{DBLP:conf/icfem/PlatzerQ09}.
We consider the central model of this study and apply \AlgoName to automate design.
It is listed as the running example \rref{model:etcs}.

\subsection{Sled}
\label{app:sled}
\begin{model}[htb]
  \setcounter{modelline}{0}
  \caption{Sled must swerve to avoid an obstacle.}\label{model:sled}
  \begin{spreadlines}{0.3ex}
  \begin{align*}
    \text{\assum}  &\left|
    \begin{aligned}
    &\mline{line:merge-assums-2} \highlightSynth{I} \land T_x > 0 \land T_y > 0 \land V>0 \land T>0 \limply [\{ \\
    \end{aligned}
    \right.\\
    \text{\ctrl} &\left|
    \begin{aligned}
      &\mline{line:merge-away} \qquad ( \ptest{\, \highlightSynth{G_1}} \seq \pumod{v_x}{-V} \; \cup \; {\ptest{\, \highlightSynth{G_2}} \seq \pumod{v_x}{V}} )\seq \\
    \end{aligned}
    \right.\\
    \text{\plant} &\,\big|\mline{line:merge-plant} \qquad (\pumod{t}{0} \seq \{x'= v_x, y' = V, t'=1 \ \& \ t \leq T \}) \\
    \text{\safe} &\,\big|\mline{line:merge-safety} \}^\ast] (y<T_y \lor x<-T_x \lor T_x<x)
  \end{align*}
  Where $I$, $G_1$ and $G_2$ are to be synthesized.
  \end{spreadlines}
  \end{model}
  This benchmark displays \AlgoName's ability to reason about state-dependent fallbacks.
  The slope of a hill is pushing a sled forward along the \(y\) axis with constant speed $v_y$.
  However, there is a wall blocking the way.
  It starts at $y$ axis position $T_y$ and extends along the $x$ axis from $-T_x$ to $T_x$.
  The sled must swerve to avoid the obstacle.
  It can either go left (with velocity $-V$) or right (with velocity $V$).
  Which action is best depends on where the sled already is.
  Swerving left is a safe strategy when the sled can pass from the $-x$ side, mathematically, $T_y>T_x+x+y$.
  Likewise, swerving right is a safe strategy when $T_y+x>T_x+y$.
  Neither action alone gives the optimal invariant, but \AlgoName's $\Inth{0}$ characterization correctly captures the disjunction to find it.
  In synthesizing the guards for the actions, \AlgoName also identifies when it is still safe to switch strategies.

  \AlgoName finds the solution below. \anntnRemark
  \begin{equation*}
    I \equiv y < T_y \land (\overbracket{T_y>T_x+x+y}^{\text{can swerve left}} \lor \overbracket{T_y+x>T_x+y}^{\text{can swerve right}}) \lor \overbracket{T_x < x \lor T_x+x < 0}^{\text{already safe}}
  \end{equation*}
  The guard for going left is
  \begin{equation*}
    \begin{split}
    G_1 \equiv & \overbracket{T_x+2TV+ y < T_y+x \land T_x+T_y>x+y \land x>=0}^{\text{despite going left one time period, can still pass from right}} \\
      \lor & \overbracket{T_x+x+y < T_y \land \neg T_x+x=0}^{\text{can pass from left}}
      \lor \overbracket{0 = T_x+x}^{\text{at the left boundary}} \\
      \lor & \overbracket{T_x+TV < x \land T_x+T_y \leq x+y}^{\text{far enough right to stay right after this cycle}} 
      \lor \overbracket{T_x+x < 0}^{\text{already left}} \\
    \end{split}
  \end{equation*}
  The guard for going right is
  \begin{equation*}
    \begin{split}
      G_2 \equiv  & \overbracket{x < 0 \land T_x + 2TV+x+y <  T_y \land  T_x + T_y +x>y}^{\text{despite going right one time period, can still pass from left}}
      \lor  \overbracket{T_x  < x}^{\text{already right}} \\
      \lor  &\overbracket{T_x +y <  T_y +x \land  T_x >x}^{\text{can pass from right}}
      \lor  \overbracket{T_x +TV+x < 0 \land  T_x + T_y +x \leq y}^{\text{far enough left to stay right after this cycle}} \\
      \lor  & \overbracket{T_y >y \land  T_x  \leq x}^{\text{right at the right boundary but there is time still to swerve}}
    \end{split}
  \end{equation*}
  There are some redundancies in these expressions, but they are correct, comprehensible, and can be simplified further.

\subsection{Parachute}
\label{app:parachute}
  \begin{model}[htb]
    \setcounter{modelline}{0}
    \caption{Parachute benchmark from \cite{DBLP:conf/itp/FultonMBP17}.}\label{model:parachute}
    \begin{spreadlines}{0.3ex}
    \begin{align*}
    \text{\assum} &\,\big|\mline{line:parachute-assums} ({T>0 \land p>0 \land g>0 \land r>0 \land m>0 \land v>0 \land \highlightSynth{I}}) \limply \{  \\
    \text{\ctrl} &\,\left|
    \begin{aligned}
        & \mline{line:parachute-ctrl-dive} \quad ({\ptest{\highlightSynth{G_1}} \seq \text{\skp}}) \cup
        ({ \ptest{\highlightSynth{G_2}} \seq \pumod{r}{p}}) \seq \\
    \end{aligned}
    \right. \\
    \text{\plant} &\,\left|
    \begin{aligned}
        &\mline{line:parachute-dyn-1} \quad (\pumod{t}{0} \seq \{x'=-v, v'=-r\cdot v^2 + g, t'=1 \ \&\ v>0 \land t \leq T \}) \\
    \end{aligned}
    \right.\\
    \text{\safe} &\,\big|\mline{line:parachute-safety} \}{^\ast}] (x \geq 0 \limply v<m)\\[-5ex]
    \end{align*}
    Where $I$, $G_1$ and $G_2$ are to be synthesized.
    \end{spreadlines}
    \end{model}
The parachute benchmark presents the challenge of dynamics with a solution that departs from the decidable fragment of real arithmetic.
A person is free falling.
At most once, they are allowed to take the action of opening a parachute.
Once they do, their air resistance changes to $p$.
The objective is to land at a speed no greater than $m$.
The benchmark is inspired by one that appears in the literature (running example in \cite{DBLP:conf/itp/FultonMBP17}).
\AlgoName uses Pegasus's Darboux Polynomial generation to solve the problem.

\AlgoName finds the solution below. \anntnRemark
\begin{equation*}
  \begin{split}
  I \equiv & \overbracket{g \geq rv^2 \land m>(gr^{- 1})^{1/2}}^{\text{either start below terminal velocity and terminal velocity without parachute is already safe}} \\
   \lor & \overbracket{m>(gp^{- 1})^{1/2} \land pv^2 \leq g}^{\text{or air terminal velocity with parachute is safe and start below terminal velocity}}
  \end{split}
\end{equation*}
The guard for not opening the parachute is that already terminal velocity without the parachute is safe, and the person has not exceeded terminal velocity yet.
\begin{equation*}
  G_1\equiv m>(gr^{- 1})^{1/2} \land rv^2 \leq g
\end{equation*}
Likewise, the guard for opening the parachute is that terminal velocity with the parachute is safe.
\begin{equation*}
  G_2\equiv m>(gp^{- 1})^{1/2} \land pv^2 \leq g
\end{equation*}

\subsection{Intersection}\label{app:intersection}
\begin{model}[htb]
  \setcounter{modelline}{0}
  \caption{Intersection benchmark.}\label{model:intersection}
  \begin{spreadlines}{0.3ex}
  \begin{align*}
    \text{\assum} &\,\big|\mline{line:intersection-assums} (B>0 \land T>0 \land v\geq 0 \land \highlightSynth{I}) \limply \{  \\
    \text{\ctrl} &\,\left|
    \begin{aligned}
        & \mline{line:intersection-ctrl} \quad ((\ptest{\highlightSynth{G_1}} \seq \pumod{a}{0}) \cup (\ptest{\highlightSynth{G_2}} \seq \pumod{a}{-B}))\seq \\
    \end{aligned}
    \right. \\
    \text{\plant} &\,\left|
    \begin{aligned}
        &\mline{line:intersection-dyn-1} \quad (\pumod{t}{0} \seq \{x'=v, v'=a, \textit{timeToRed}'=-1, t'=1 \ \&\ t\leq T \land v\geq 0 \}) \\
    \end{aligned}
    \right.\\
    \text{\safe} &\,\big|\mline{line:intersection-safety} \}{^\ast}] ((\textit{timeToRed}>0 \land v\neq 0) \lor \neg(x=0))\\[-5ex]
    \end{align*}
    Where $I$, $G_1$ and $G_2$ are to be synthesized.
    \end{spreadlines}
    \end{model}
  The intersection benchmark is a simple example of a system with free choice and state dependent fallback.
  A car sees a yellow light, and must decide whether to coast past the intersection, or to stop before it.
  It may never stop at the intersection, which is located at $x=0$.
  Whether it would be safe to stop or to coast depends on the car's current position and velocity.

  \AlgoName finds the solution below. \anntnRemark
  \begin{equation*}
    \begin{split}
    I &\equiv \overbracket{x>0}^{\text{safe if already past the intersection}}\\
    & \lor \overbracket{x < 0 \land (v\leq 0 \lor v(\textit{timeToRed}\cdot v+x)>0)}^{\text{otherwise, before the intersection, safe if velocity is already 0 or could coast past intersection}}\\
    & \lor {v>0 \land \big(timeToRed>0 \land (}\\
    & \phantom{ \lor} \phantom{v>0 \land (} \overbracket{B < \frac{v}{\textit{timeToRed}} \land B\cdot \textit{timeToRed}^2 < 2(\textit{timeToRed}\cdot v+x)}^{\text{signal flips before car stops. Even braking, car crosses intersection before signal flips}}\\
    & \phantom{ \lor} \phantom{v>0 \land (} \lor \overbracket{v^2+2\cdot B\cdot x\neq0 \land B\geq \frac{v}{\textit{timeToRed}})}^{\text{stop somewhere that's not the intersection, the signal flips after the car stops}}\\
    \end{split}
  \end{equation*}

  The guard for coasting has many repeated clauses, so we first explain them before presenting the expression.
  Assuming $v$ positive and $x\leq 0$,
  $C \equiv v^3+2Bv(Tv+x) < 0$ means that after one time period of coasting, the car still stops before the intersection.
  $D_1 \equiv 0=3\textit{timeToRed}+2v^{- 1}x$ means that the signal flips when the car is $2/3$rds along the way of coasting to the intersection.
  $D_2 \equiv v(3\textit{timeToRed}\cdot v+2x)>0$ means that the signal flips after the car is $2/3$rds along the way of coasting to the intersection.
  $D_3 \equiv v(3\textit{timeToRed}\cdot v+2x) < 0$ means that the signal flips before the car is $2/3$rds along the way of coasting to the intersection.
  $E \equiv B=\textit{timeToRed}^{- 1}v$ means that the signal flips exactly when the car halts if it starts braking now.
  $F \equiv 0=\textit{timeToRed}+v^{- 1}x$ means that the signal flips exactly when the car reaches the intersection by coasting.
  $G \equiv 2B+v^2(\textit{timeToRed}\cdot v+x)^{- 1}=0$ means that the car will be at the intersection when the signal flips if it starts braking now.
  $H_1 \equiv v(\textit{timeToRed}\cdot v+x) < 0$ means that if the car coasts, it will be before the intersection when the signal flips.
  $H_2 \equiv v(\textit{timeToRed}\cdot v+x) > 0$ means that if the car coasts, it will be after the intersection when the signal flips.
  \begin{equation*}
    \begin{split}
    G_1 \equiv
    & \overbracket{v=0}^{\text{already still and safe}} \lor (v>0 \land ( \\
    & \phantom{\land} \overbracket{x>0}^{\text{already past intersection}} \lor x<=0 \land ( \\
    & \phantom{\land\lor} D_1 \land (\overbracket{ (T < timeToRed \land E) \lor (C \land \neg E) }^{\text{can stop before intersection after cycle of coasting}}) \\
    & \phantom{\land} \lor \overbracket{C \land ( F \lor \neg G \land (D_3 \lor D_2 \land  H_1))}^{\text{can stop before intersection after cycle of coasting (but $D_1$ doesn't have to hold)}} \\
    & \phantom{\land} \lor  \overbracket{(D_2 \land H_1 \lor D_3) \land T < timeToRed \land G}^{\text{after one cycle of coasting and then braking, car passes intersection}} \\
    & \phantom{\land} \lor  \overbracket{H_2}^{\text{can coast past intersection before signal switches}})))
    \end{split}
  \end{equation*}

  Likewise, the guard for braking has many repeated clauses.
  $P \equiv 2B+v^2/x \leq 0$ means that the car won't stop before the intersection.
  $Q \equiv \textit{timeToRed}>\frac{BT^2+2x}{2(BT- v)}$ means that even after one braking cycle, the car can cross the intersection by coasting.
  $R \equiv x(v^2+2Bx)>0$ means that the car will stop before the intersection.
  $S \equiv B\cdot \textit{timeToRed}+(v^2+2Bx)^{1/2}>v$ means that the car will stop before the intersection before the signal turns red.
  $U \equiv 2(Tv+x) < BT^2$ means that the car will stop before time T.
  $V \equiv BT^2=2(Tv+x)$ means that if it were to brake, the car will come to a stop at time T.
  $W \equiv BT^2 < 2(Tv+x)$ means that if it were to brake, the car will come to a stop after time T.
  Because of structural similarities with $G_1$, we do not provide a full annotation.
  \begin{equation*}
    \begin{split}
    G_2 \equiv
    &x<0 \land (
      v(Tv+x)<0 \land (P \land Q \lor R) \\
      &\lor v(Tv+x)>0 \land T < -2v^{-1}x \land ( W \land S \lor P \land Q \land U \\
      & \phantom{\land } \lor R \lor V \land \textit{timeToRed}>T) \\
      &\lor 0=T+v^{-1}x \land ( P \land Q \land U \lor R ) \\
      &\lor T \geq -2v^{-1}x \land ( R \lor S \land x(v^2+2Bx)<0))
    \lor x \geq 0
    \end{split}
  \end{equation*}

\subsection{Curvebot}
\label{app:curvebot}
\begin{model}[htb]
  \setcounter{modelline}{0}
  \caption{Curvebot benchmark.}\label{model:curvebot}
  \begin{spreadlines}{0.3ex}
  \begin{align*}
    \text{\assum} &\,\big|\mline{line:curvebot-assums} (T>0 \land \highlightSynth{I} ) \limply \{  \\
    \text{\ctrl} &\,\left|
    \begin{aligned}
        & \mline{line:curvebot-ctrl} \quad ((\highlightSynth{G_1} \seq \pumod{\textit{om}}{1}) \cup (\highlightSynth{G_2} \seq \pumod{\textit{om}}{0}) \cup (\highlightSynth{G_3} \seq \pumod{om}{-1})) \seq \\
    \end{aligned}
    \right. \\
    \text{\plant} &\,\left|
    \begin{aligned}
        &\mline{line:curvebot-dyn-1} \quad (\pumod{t}{0} \seq \{x'=v,y'=w,v'=\textit{om}\cdot w,w'=-\textit{om}\cdot v, t'=1 \ \&\ t\leq T \}) \\
    \end{aligned}
    \right.\\
    \text{\safe} &\,\big|\mline{line:curvebot-safety} \}{^\ast}] \neg(x=0 \land y=0)\\[-5ex]
    \end{align*}
    Where $I$, $G_1$, $G_2$ and $G_3$ are to be synthesized.
    \end{spreadlines}
    \end{model}
Curvebot models a Dubin's car that must avoid an obstacle at (0,0).
The dynamics result in a solution that is not in the decidable fragment of arithmetic, so \AlgoName again uses Pegasus to find a controllable invariant.

Our implementation generates the optimal invariant, consisting of everywhere except the origin.
\anntnRemark
\begin{equation*}
  \begin{split}
  I \equiv \neg x=0 \lor \neg y=0
  \end{split}
\end{equation*}
The guard for setting $om$ to 1 is simply that the origin does not lie on the resulting circular path.
\begin{equation*}
  G_1 \equiv y=0 \land \neg 2w+x=0 \lor \neg y(2wx+x^2+y((- 2)v+y))=0
\end{equation*}
The guard for setting $om$ to 0 is that the origin does not lie in the straight line segment of length $T(v^2+w^2)^(1/2)$ ahead.
\begin{equation*}
  \begin{split}
  G_2 \equiv & \overbracket{v<0 \land (\neg y=0 \land (0=w \lor 0=x)}^{\text{going in the -$x$ direction $\cdots$}} \\
  & \phantom{\lor} \overbracket{\lor x>0 \land (\neg w=0 \land \neg y=wx/v \lor v(Tv+x)<0) \lor x<0)}^{\text{$\cdots$ is safe}} \\
  \lor & \overbracket{v=0 \land (\neg x=0 \lor y>0 \land (w\geq 0 \lor w(Tw+y)<0)}^{\text{purely y motion $\cdots$}} \\
  & \phantom{\lor} \lor \overbracket{y<0 \land (w(Tw+y)<0 \lor w\leq 0))}^{\text{$\cdots$ is safe}} \\
  \lor & \overbracket{v>0 \land (\neg y=0 \land (0=w \lor 0=x)}^{\text{going in the $x$ direction $\cdots$}} \\
  & \phantom{\lor} \lor \overbracket{x>0 \lor (\neg w=0 \land \neg y=wx/v \lor v(Tv+x)<0) \land x<0)}^{\text{$\cdots$ is safe}} \\
  \end{split}
\end{equation*}
Like $G_1$, the guard for setting $om$ to -1 is that the origin does not lie on the resulting circular path.
\begin{equation*}
  G_3 \equiv y=0 \land \neg 2w=x \lor \neg y(x^2+y(2v+y))=2wxy
\end{equation*}
This solution is almost optimal.
It only misses the cases for $G_1$ and $G_2$ where despite the obstacle lying on the circular path, $T$ is small enough that there is time to switch paths before collision.

\subsection{Corridor}
\label{app:corridor}
Corridor, shown in \rref{model:corridor}, is an example of a system requiring unrolling.

\subsection{Power Station}
\label{app:power-station}
\begin{model}[htb]
  \setcounter{modelline}{0}
  \caption{Power Station benchmark.}\label{model:power-station}
  \begin{spreadlines}{0.3ex}
  \begin{align*}
    \text{\assum} &\,\left|
    \begin{aligned}
      &\mline{line:power-station-assums-1} (T=1 \land i\geq 0 \land \textit{chargeRate}=7000000 \land J=100 \\
      &\mline{line:power-station-assums-2} \land V=2000 \land R=5 \land \textit{quota}=3000 \land \highlightSynth{I}) \limply [\{  \\
    \end{aligned}
    \right. \\
    \text{\ctrl} &\,\left|
    \begin{aligned}
        & \mline{line:power-station-charge} \quad ((\highlightSynth{G_1} \seq \pumod{i}{0} \seq \pumod{\textit{slope}}{\textit{chargeRate}}) \\
        & \mline{line:power-station-discharge} \quad \cup (\highlightSynth{G_2} \seq \pumod{i}{J} \seq \pumod{\textit{slope}}{-i\cdot V})) \seq \\
    \end{aligned}
    \right. \\
    \text{\plant} &\,\left|
    \begin{aligned}
        &\mline{line:power-station-dyn-1} \quad (\pumod{t}{0} \seq \{\textit{produced}'=i\cdot V - i^2\cdot R,\\
        &\mline{line:power-station-dyn-2} \textit{stored}'=\textit{slope}, t'=1, \textit{gt}'=-1 \ \&\ t\leq T \}) \\
    \end{aligned}
    \right.\\
    \text{\safe} &\,\big|\mline{line:power-station-safety} \}{^\ast}] (\textit{stored}>0 \land (\textit{gt}>0 \lor \textit{produced}>\textit{quota}))\\[-5ex]
    \end{align*}
    Where $I$, $G_1$ and $G_2$ are to be synthesized.
    \end{spreadlines}
    \end{model}
  Power station is an example of a system that needs two steps of unrolling in order to reach the optimal invariant.
  A power station capable of producing $7000$ kW can choose between charging (\rref{line:power-station-charge}) and distributing out stored power at current $100$ A and voltage of $2000$ V (\rref{line:power-station-discharge}).
  Its objective is to meet an energy quota of $3000$ J, excluding loss due to resistance of $5\Omega$ by the time that timer $gt$ counts down to 0.
  The station must never reach a state where it has no stored power left.
  The system is modeled in \rref{model:power-station}.
  The zero-shot invariant corresponds to the case where the station has already met its quota and is now charging.
  Discharge is not included in the invariant because regardless of how high stored energy is, in the unbounded time of the zero-shot invariant, a station that has chosen to distribute power will eventually run out of it, thus violating the condition $stored>0$.
  The one shot invariant catches the case where the station first chooses to discharge till it meets its quota, and then flips to charging for infinite time.
  The two shot invariant, which is finally optimal, catches the case where the station first chooses to charge till it has enough energy stored to meet its quota, then discharges, and finally switches back to charging mode for infinite time.
  \AlgoName finds the solution below. \anntnRemark
  \begin{equation*}
    \begin{split}
      I &\equiv \textit{produced}\leq 3000 \land \overbracket{(3000(-51+50 \textit{gt})+ \textit{produced}>0}^{\text{enough time to charge $\cdots$}}\\
      & \overbracket{ \land 7000000 \textit{gt} + 48 \textit{produced} + \textit{stored} > 7344000 \land 437500 \textit{gt} + 3 \textit{produced} < 459000}^{\text{$\cdots$ then distribute}} \\
      \lor & \overbracket{150000 \textit{gt} + \textit{produced} > 3000 \land 4 \textit{produced} + 3 \textit{stored} > 612000)}^{\text{enough stored power to distribute and meet quota}} \\
      \lor & \overbracket{stored>0 \land (437500 \textit{gt} + 3 \textit{produced} \geq 459000 \lor \textit{produced} > 3000)}^{\text{without drawing stored energy, enough time to charge then distribute}} \\
    \end{split}
  \end{equation*}
  The guard for charging checks that the choice to charge for a cycle still leaves enough time for distributing the power.
\begin{equation*}
  \begin{split}
    G_1 \equiv & \overbracket{\textit{gt}>1 \land 3000(\textit{gt}\cdot 50-51)+\textit{produced}>0}^{\text{enough time to charge then distribute}} \\
    & \lor \overbracket{\textit{gt}\cdot 35 < 36 \land \textit{produced}>3000}^{\text{quota already met}} \\
  \end{split}
\end{equation*}
The guard for distributing basically checks that the choice to discharge for a cycle still leaves enough stored energy.
\begin{equation*}
  \begin{split}
    G_2 \equiv & \overbracket{\textit{produced}\leq 3000 \land 3000(50 \textit{gt}-51) + \textit{produced}\leq 0}^{\text{must distribute to stay safe}} \\
    & \lor \overbracket{3000(50 \textit{gt}-51) + \textit{produced}>0 \land stored>200000 \land \textit{gt}>1}^{\text{enough stored energy to distribute for a cycle, can produce enough afterwards}} \\
    & \lor \overbracket{\textit{gt}\leq 1 \land stored>200000 \land \textit{produced}>3000}^{\text{enough stored energy for a cycle, quota satisfied}} \\
  \end{split}
\end{equation*}

\subsection{Coolant}
\label{app:coolant}
\begin{model}[htb]
  \setcounter{modelline}{0}
  \caption{Coolant benchmark.}\label{model:coolant}
  \begin{spreadlines}{0.3ex}
  \begin{align*}
    \text{\assum} &\,\left|
    \begin{aligned}
      & \mline{line:coolant-assums-1} (T>0 \land F>0 \land \textit{minAbsorbed}>0 \land \textit{maxDischarge}>0\\
      & \mline{line:coolant-assums-2} \land \textit{tempDiff}>0 \land c>0 \land \highlightSynth{I}) \limply \{  \\
    \end{aligned}
    \right. \\
    \text{\ctrl} &\,\left|
    \begin{aligned}
        & \mline{line:coolant-charge} \quad ((\highlightSynth{G_1} \seq \pumod{f}{0}) \\
        & \mline{line:coolant-discharge} \quad \cup (\highlightSynth{G_2} \seq \pumod{f}{F})) \seq \\
    \end{aligned}
    \right. \\
    \text{\plant} &\,\left|
    \begin{aligned}
        &\mline{line:coolant-dyn-1} \quad (\pumod{t}{0} \seq \{\textit{absorbed}'=f\cdot c\cdot \textit{tempDiff},\\
        &\mline{line:coolant-dyn-2} \textit{discharged}'=f, t'=1, \textit{gt}'=-1 \ \&\ t\leq T \}) \\
    \end{aligned}
    \right.\\
    \text{\safe} &\,\big|\mline{line:coolant-safety} \}{^\ast}] (\textit{discharged}<\textit{maxDischarge} \land (\textit{gt}>0 \lor \textit{absorbed}\geq \textit{minAbsorbed}))\\[-5ex]
    \end{align*}
    Where $I$, $G_1$ and $G_2$ are to be synthesized.
    \end{spreadlines}
    \end{model}
  The coolant benchmark is an example of a system that requires unrolling.
  A coolant system draws water at either a rate of $F$ or 0 $m^3/s$ from a reservoir.
  It runs the water through a heat exchanger, drawing heat proportional to the specific heat of water $c$ and the temperature difference between the water and the system.
  It then discharges the water into a river at the rate that it draws water.
  The cooling system must absorb at least $minAbsorbed$ heat by the time that timer $gt$ counts down to 0.
  But it must also not discharge more that $maxDischarged$ water into the river to avoid environmental damage.
  The benchmark is modeled in \rref{model:coolant}.

  \AlgoName finds the solution below. \anntnRemark
 \begin{equation*}
  \begin{split}
    I &\equiv \overbracket{\textit{absorbed}\geq \textit{minAbsorbed} \land \textit{discharged}<\textit{maxDischarge}}^{\text{already absorbed enough heat}} \\
    \lor & \overbracket{\textit{gt}>0 \land (\textit{discharged}+F\cdot T < \textit{maxDischarge}}^{\text{drawing water one cycle will not exceed discharge limit}}\\
    & \phantom{\lor} \overbracket{\land \textit{maxDischarge}\leq \textit{discharged}+F\cdot (\textit{gt}+T)}^{\text{ but drawing water till timer $gt$ runs out will}} \\
    & \phantom{\lor} \overbracket{\land c>(\textit{absorbed}-\textit{minAbsorbed})\cdot \textit{tempDiff}}^{\text{drawing as much water as discharge limit allows $\cdots$}} \\
    & \phantom{\lor -} \overbracket{\cdot (\textit{discharged} - \textit{maxDischarge} + F\cdot T)^{- 1}}^{\text{$\cdots$ absorbs enough heat}} \\
    \lor & \overbracket{ \textit{discharged}+F\cdot (\textit{gt}+T) < \textit{maxDischarge} }^{\text{drawing water now $\cdots$}}\\
    & \phantom{\lor} \overbracket{ \land \textit{gt}^{- 1}\cdot (\textit{absorbed}-\textit{minAbsorbed}+c\cdot F\cdot \textit{gt}\cdot \textit{tempDiff})\geq 0)}^{\text{$\cdots$ till the timer runs out is safe}} \\
  \end{split}
\end{equation*}
The guard for drawing no water just needs to check that it still has enough time to absorb enough heat.
\begin{equation*}
  \begin{split}
    G_1 \equiv & \overbracket{\textit{absorbed}\geq \textit{minAbsorbed}}^{\text{already absorbed enough heat}} \\
    \lor & \overbracket{\textit{gt}>T \land ( (F=(\textit{maxDischarge}-\textit{discharged})\cdot \textit{gt}^{- 1}}^{\text{drawing water will $\cdots$}} \\
    & \phantom{\lor} \overbracket{\lor \textit{discharged}+T\cdot F\geq \textit{maxDischarge}}^{\text{$\cdots$ exceed the $\cdots$}} \\
    & \phantom{\lor} \overbracket{\lor \textit{gt}\cdot (\textit{discharged}+\textit{gt}\cdot F-\textit{maxDischarge})>0)}^{\text{$\cdots$ discharge limit}} \\
    \lor & \overbracket{\textit{gt}\cdot (\textit{discharged}+\textit{gt}\cdot F-\textit{maxDischarge}) < 0}^{\text{despite deferring one cycle $\cdots$}} \\
    & \phantom{\lor} \overbracket{\land (\textit{absorbed}-\textit{minAbsorbed})\cdot (\textit{gt}-T)^{- 1}+F\cdot (c\cdot \textit{tempDiff})\geq 0)}^{\text{$\cdots$ can still absorb enough heat}} \\
  \end{split}
\end{equation*}
The guard for drawing water checks that drawing water for one cycle will not exceed the discharge limit.
\begin{equation*}
  \begin{split}
    G_2 \equiv & \textit{discharged}+F\cdot T < \textit{maxDischarge}
  \end{split}
\end{equation*}

\clearpage

\section{Simplification Impact Example}\label{app:reduce-example}

Reduction of even a loop free \dGL formula, if done na\"ively, can produce very large expressions.
Not only are large expression hard to reason about, but they also slow down quantifier elimination.
The effect on QE is so stark that the benchmark \texttt{Curvebot}, which completes in 26s with our simplification stack, times out after 20 minutes without simplification.
Our simplification uses Mathematica's \texttt{Simplify} function.
Additionally, to catch common patterns that \AlgoName produces but Mathematica does not handle, we implemented a custom simplifier with \texttt{Egg} and \texttt{z3}.

The impact of simplification can be dramatic.
For example, the $I_0$ invariant produced by \AlgoName for \texttt{Curvebot} \textit{without} simplification is the following expression:
\begin{equation*}
  \begin{split}
    &((v \neq wx/y \land y\neq 0 \lor y=0 \land (w=0 \land (v\geq 0 \land x>0 \lor v\leq 0 \land x < 0)\\
    & \lor w\neq 0 \land x\neq 0) \lor w\geq 0 \land y>0 \lor w\leq 0 \land y < 0) \\
    & \lor x=0 \land 2v\neq y \land y\neq 0 \lor x\neq 0 \land (2w+x\neq 0 \lor y\neq 0) \\
    & \land (y=0 \lor 2vx\cdot y\neq x\cdot (2wx+x^2+y^2))) \lor (2w\neq x \lor y\neq 0) \\
    & \land x\neq 0 \land (y=0 \lor x^3+x\cdot y\cdot (2v+y)\neq 2wx^2) \\
    & \lor x=0 \land 2v+y\neq 0 \land y\neq 0
  \end{split}
\end{equation*}
Running this formula through our simplifier produces the expression $y\neq 0 \lor (y=0 \land 0\neq x)$.
Thus, the expression is actually just $\neg (x=0\land y=0)$.

Besides being much easier to read, simpler expressions are also much faster to eliminate quantifiers from.
Since doubly exponential quantifier elimination is the bottleneck in scaling to increasingly complex problems, simplification is crucial to the scalability of \AlgoName.
In the case of our example above, without simplification, \texttt{Curvebot} gets stuck on the following QE problem.
\begin{equation*}
  \begin{split}
    &\forall v_1,w_1,x_1,y_1 (((((v^2+w^2=v_1^2+w_1^2 \land vw+vx-wy-xy \\
    & =v_1w_1+v_1x_1-w_1y_1-x_1y_1) \land w^2/2+v y-y^2/2=w_1^2/2+v_1y_1-y_1^2/2) \\
    & \land v-y=v_1-y_1) \land w^2/2+w x+x^2/2=w_1^2/2+w_1x_1+x_1^2/2)\\
    & \land w+x=w_1+x_1 \limply (((v_1\neq w_1x_1/y_1 \land y_1\neq 0 \\
    & \lor y_1=0 \land (w_1=0 \land (v_1\geq 0 \land x_1>0 \lor v_1\leq 0 \land x_1 < 0)\\
    & \lor w_1\neq 0 \land x_1\neq 0) \lor w_1\geq 0 \land y_1>0 \lor w_1\leq 0 \land y_1 < 0) \\
    & \lor x_1=0 \land 2v_1\neq y_1 \land y_1\neq 0 \lor x_1\neq 0 \land (2w_1+x_1\neq 0 \\
    & \lor y_1\neq 0) \land (y_1=0 \lor 2v_1x_1\cdot y_1\neq x_1\cdot (2w_1x_1+x_1^2+y_1^2))) \\
    & \lor (2w_1\neq x_1 \lor y_1\neq 0) \land x_1\neq 0 \land (y_1=0 \\
    & \lor x_1^3+x_1\cdot y_1\cdot (2v_1+y_1)\neq 2w_1x_1^2) \\
    & \lor x_1=0 \land 2v_1+y_1\neq 0 \land y_1\neq 0) \land \neg (x_1=0 \land y_1=0))
  \end{split}
\end{equation*}
But with simplification applied regularly, instead the corresponding call to QE is
\begin{equation*}
  \begin{split}
  &\forall v_1,w_1,x_1,y_1 (((((v^2+w^2=v_1^2+w_1^2 \land vw+vx-wy-xy \\
  & =v_1w_1+v_1x_1-w_1y_1-x_1y_1) \land w^2/2+v y-y^2/2=w_1^2/2+v_1y_1-y_1^2/2) \\
  & \land v-y=v_1-y_1) \land w^2/2+w x+x^2/2=w_1^2/2+w_1x_1+x_1^2/2)\\
  & \land w+x=w_1+x_1 \limply \neg (x_1=0 \land y_1=0))
  \end{split}
\end{equation*}
which terminates in 70 milliseconds instead of timing out.

Finally, even on benchmark examples that would terminate without the use of \emph{eager simplification}, enabling this optimization often results in shorter solutions. \rref{tab:simplification-impact} shows the percentage reduction in size of solutions due to simplification, where size is measured in terms of number of characters. A $0\%$ reduction means no change and a $50\%$ reduction means cutting formula-size in half.

\begin{table}
  \caption{Simplification impact on solution size.}
  \label{tab:simplification-impact}
  \centering
  \begin{tabular}{|l|r|}
    \hline
    \textbf{Benchmark} & \textbf{\% Solution Size Reduction}  \\
    \hline
    ETCS Train & 62\% \\
    Sled & 41\% \\
    Intersection & 38\% \\
    Curvebot & $\infty$ \\
    Parachute & 5\% \\
    Coolant & 92\% \\
    Corridor & 85\% \\
    Power Station & 55\% \\
    \hline
  \end{tabular}
\end{table}
} 

\end{document}